\documentclass[journal ]{new-aiaa}
\usepackage[utf8]{inputenc}
\usepackage{textcomp}

\usepackage{graphicx}
\usepackage{amsmath}
\usepackage[version=4]{mhchem}
\usepackage{siunitx}
\usepackage{longtable,tabularx}
\usepackage{comment}
\usepackage[super]{nth}
\newcommand\norm[1]{\lVert#1\rVert}

\usepackage{hyperref}

\title{Loss of Control Prevention of an Agile Aircraft: Dynamic Command Saturation Approach}

\author{Ege Ç. Altunkaya \footnote{Research Assistant, Aviation Institute, Aerospace Research Center, Istanbul Technical University, Istanbul, Türkiye, 34469; altunkaya16@itu.edu.tr} and Akın Çatak \footnote{Research Assistant, Aviation Institute, Aerospace Research Center, Istanbul Technical University, Istanbul, Türkiye, 34469.}}
\affil{Aerospace Research Center, Istanbul Technical University, Istanbul, Türkiye, 34469}
\author{Emre Koyuncu \footnote{Assoc. Prof., Aerospace Research Center, Istanbul Technical University, Istanbul, Türkiye, 34469.}}
\affil{Aerospace Research Center, Istanbul Technical University, Istanbul, Türkiye, 34469}
\author{İbrahim Özkol \footnote{Prof., Aerospace Research Center, Istanbul Technical University, Istanbul, Türkiye, 34469.}}
\affil{Aerospace Research Center, Istanbul Technical University, Istanbul, Türkiye, 34469}

\begin{document}

\maketitle

%%

%%

\begin{comment}

\section*{Nomenclature}

{\renewcommand\arraystretch{1.0}
\noindent\begin{longtable*}{@{}l @{\quad=\quad} l@{}}
$\alpha$, $\beta$ & Angle of attack and sideslip, deg \\
$\phi$, $\theta$, $\psi$ & Bank, pitch, and yaw angle, deg \\
$\delta_{HT_L}$, $\delta_{HT_R}$ & Left and right horizontal tail deflection, deg \\
$\delta_{A_L}$, $\delta_{A_R}$ & Left and right aileron deflection, deg \\
$\delta_{R}$ & Rudder deflection, deg \\
$\tau_c$ & Control moment coefficient vector \\
$\Phi$ & Control effectiveness matrix \\
$p, q, r$ & Roll, pitch, and yaw rate, deg/s \\
$C_{l}$, $C_{m}$, $C_{n}$ & Rolling, pitching, and yawing moment coefficient \\
$C_{y_r}$ & Y-axis force coefficient w.r.t yaw rate, 1/rad \\
$C_{D_q}$ & Drag force coefficient w.r.t pitch rate, 1/rad \\
$C_{L_q}$ & Lift force coefficient w.r.t pitch rate, 1/rad \\
$\Bar{c}$, $b$ & Mean aerodynamic chord and wing span, m \\
$J$ & Inertia tensor, $kgm^2$ \\
$S$ & Wing area, $m^2$ \\
$\mathbb{A_M}, \Delta \mathbb{A_M}$ & Attainable and incremental attainable moment set \\
\end{longtable*}}

\end{comment}

%%

\section{Introduction}
\lettrine{T}{he} loss of control, often resulting from compromised stability and/or controllability, jeopardizes flight safety and is a leading factor in fatal crashes \cite{ms7}. Numerous factors contribute to this loss of control, including adverse onboard conditions, such as system malfunctions or pilot error, and unexpected vehicle upsets caused by environmental disturbances or mechanical failures \cite{statistics1, statistics2}. The diverse array of reasons and their interactions inherently elevate the risk of control loss, as extensively outlined in terms of fatal crash statistics linked to the control loss in \cite{statistics1} and \cite{statistics2}. Moreover, a detailed analysis of the causes behind the control loss reveals that the pilot-induced loss of control, a significant contributing factor, involves abrupt and aggressive maneuvers that can trigger a destabilized flight \cite{statistics2}. Despite the desirability of performing aggressive maneuvers without compromising safety, the ability to do so remains constrained by the aircraft’s aerodynamic characteristics, controllability, and flight dynamics. Additionally, according to \cite{ms4}, reasons for loss of control can be categorized under five major headings: (1) stall, (2) sideslip-induced rolls, (3) rolls from other causes\textemdash such as wake vortices, structural failures, and lateral pilot-induced oscillations\textemdash, (4) pilot-induced oscillations, and (5) yaw departures\textemdash often resulting from wake encounters or improper rudder input. However, the loss of control is a broad concept that cannot be fully captured by simple analyses; nevertheless, a basic definition can be stated as follows: it is "a significant deviation of the aircraft from the intended flight path or operational envelope," as defined by the Commercial Aviation Safety Team (CAST) \cite{ms3}. Moreover, CAST has proposed quantitative metrics to measure loss of control in \cite{ms2}. Thereby designing an aircraft with a nonlinear flight control law is not always sufficient and does not imply that the aircraft can perform any desired of maneuver. There are numerous restrictions that hinder an agile aircraft from conducting agile maneuvers, such as high angle of attack kinematics, inertia-coupling, and flight and/or dynamic envelope violation. The factors described above may trigger an upset condition such as loss of control, departure, spin, stall, or even compromise the aircraft structural integrity. In summary, loss of control is a flight dynamics stability and control issue where the aircraft response deviates from controllable predictable conditions, involving non-linear effects like nonlinear kinematic and inertial coupling, which cause large amplitude responses to small changes in state variables or result in bounded oscillatory or divergent dynamics. As a consequence, it often leads to significant angular rates and displacements, making it challenging for the pilot to maintain control of the aircraft \cite{statistics3}. Therefore, to prevent any upset condition, one has to delve into presumable reasons for these catastrophic consequences. 

Since loss of control still persists as a critical issue, numerous studies inherently exist to address it from various perspectives. Hereby, ensuring safe flight has been extensively studied through flight envelope protection \cite{ms21, ms20, ms17, ms16, ms15, ms14, ms13, ms11, ms10}, prevention of loss of control \cite{ms8, ms16, masterThesis}, and implementation of command/reference governor methods \cite{ms23, ms24}. Lombaerts et al. \cite{ms16} detail the design process of the adaptive algorithm and its integration into a piloted simulator for evaluation. The paper discusses the methodology employed in simulator testing, including the scenarios used to assess the algorithm's performance. The boundaries of the flight envelope are determined adaptively, and the flight envelope-aware flight control system serves to prevent loss of control. Finally, they implement the proposed architecture into a simulator environment of a commercial airline
aircraft, for further examination. Yu et al. \cite{ms11} present an approach for designing safety control systems for aircraft that incorporate flight envelope protection and reference command generation. The approach is based on a nonlinear control framework that combines feedback control with feedforward reference commands to achieve robust and safe flight, accommodating actuator faults and ensuring operation within the predetermined safe flight envelope. The results show that the approach is effective at maintaining stable flight and protecting the aircraft from operating outside of its safe flight envelope. Despite the existence of adaptive flight envelope determination methods, there also exist other online safe flight envelope estimation studies \cite{ms19, ms5} besides the ANN-based flight envelope estimation approach \cite{ms6}.

Nevertheless, various studies strongly emphasize the distinction between the flight envelope and dynamic envelope of an aircraft \cite{ms1, ms7}. In other words, the dynamic envelope defines the region within the aircraft's state space where safe controllability is assured, preventing the occurrence of events leading to loss of control. Consequently, the notion of ensuring flight conditions solely within the traditional flight envelope for the prevention of the loss of control becomes a matter of debate. Particularly, Zheng et al. \cite{ms1} study the determination of the dynamic envelope of the baseline aircraft, NASA's Generic Transport Model. They generate the dynamic envelope using differential manifold theory and claim that this method is able to capture the coupling effects more accurately. Moreover, a correlation between the dynamic envelope and control coefficients of PI controller or pilot command of pitch angle, $\theta$ is established. Finally, based on the dynamic envelope enlargement strategy, an envelope protection architecture is proposed and assessed.

On the other hand, various studies aim to estimate the maneuvering envelope through reachability analysis, such as \cite{ms9, ms18, ms8, ms5}. However, these approaches are not suitable for the real-time adaptation due to the high computational cost associated with numerically solving the Hamilton-Jacobi-Bellman equations. Alongside, the authors in \cite{reachability1} propose a real-time implementable online flight envelope detection mechanism that merges nonlinear reachability analysis with the classification capabilities of artificial neural networks. This online detection method is further utilized for accommodating structural failures in an aircraft, thereby preventing a potential loss of control, where the failure-induced loss of control is also a critical aspect and extensively addressed in \cite{edwards}. Furthermore, in \cite{ms7, ms25}, loss of control is thoroughly investigated in the sense of nonlinear dynamics, and bifurcation analysis is put forth. However, of course, bifurcation analyses can only be done in the design phase of the aircraft.

Last but not least, maneuver limiting is also addressed in various studies using the command and reference governor approaches. Simon et al. \cite{ms23} propose a method for limiting the angle of attack and load factor using the command and reference governors. The implementation has been carried out in the simulation environment used for the Jas39 Gripen fighter aircraft. In this study, both the angle of attack and load factor commands are saturated, leading to a significant reduction in overshoot in the angle of attack and load factor responses at their limit values by adjusting the command governor gain $\beta$, constrained by the corresponding flight envelope. As a consequence, since the predetermined safe flight envelope is protected, a loss of control is not observed during the desired maneuvers. A similar effort has been described in \cite{ms24} for the flight envelope protection during high-angle-of-attack maneuvers. Albeit the proposed methods are successful in achieving their primary objectives, i.e., flight envelope protection, \textit{a priori} flight envelope information is a must; otherwise, command and reference governors cannot determine the boundaries of the flight envelope in which the aircraft can be controlled without entering an upset condition.

From an industrial perspective, the primary goal is "care-free" maneuvering without compromising agility and safety, as emphasized in \cite{ms25}. However, aircraft certification remains a constant requirement in the industry, making conventional methods the preferred choice for safety. Traditional state limiters, such as angle of attack, load factor, bank angle, pitch angle, and roll rate limiters, are still in use. These limiters are scheduled based on Mach number, dynamic pressure, angle of attack, sideslip, and/or roll rate. Nevertheless, designing these state limiters for safe flight requires a labor-intensive effort. Extensive aerodynamic analyses are necessary to investigate the stability characteristics of the designed aircraft. For this purpose, analysis is typically conducted of several aircraft dynamics parameters, such as Lateral Control Departure Parameter (LCDP) and $C_{{n_\beta}_{dyn}}$. This analysis often involves varying the relevant states that define the limited envelope and using criteria to predict the aircraft's tendency to departure based on the values of both the LCDP and $C_{{n_\beta}_{dyn}}$ parameters. Careful analysis of these parameters is of significant importance to determine the departure angle of attack value, and their variations with changing flight circumstances must be considered for scheduling. Importantly, these evaluations are not limited to the early design phases but must be extended and verified through the flight tests. Consequently, these aspects contribute to the labor and cost excessively.

According to the inferences drawn from the previous research, as elaborated previously, the issue of the loss of control has been discussed within the context of the flight envelope protection. Additionally, previous studies often use pre-established flight envelope databases. As a result, the aircraft can be controlled with prior knowledge of critical values to avoid, based on the known limits of the flight envelope. However, confining the concept of the loss of control solely within the flight envelope protection does not seem inclusive, and the consideration of extreme and coupling maneuvers has been lacking in most previous research.

\subsection{Objective, Methodology, and Contributions}

This study aims to develop an online maneuver limiting strategy rather than a flight envelope protection system, with the goal of performing the most extreme maneuvers possible\textemdash beyond the limits imposed by conventional state limiters. Therefore, the goal is to perform maneuvers as extreme as possible by limiting only the pilot commands, rather than structural or attitude features such as load factors, airspeed, and angular orientations. To achieve this, a two-step strategy is proposed: (1) online loss of control detection using an incremental attainable moment set, and (2) online loss of control prevention using pilot command limiting with a Lyapunov-based command limiter. Based on these evaluations, the followings are the contributions of this study:

\begin{itemize}
    \item This study does not use pre-existing \textit{a priori} information for defining avoidance limits. Instead, it focuses on ensuring maneuver stability by intervening and limiting only the pilot commands for angle of attack ($\alpha$), roll rate ($p$), pitch rate ($q$), and yaw rate ($r$) in-flight. This process utilizes a combination of Lyapunov's direct method and inverse dynamics.

    \item The concept of incremental attainable moment set and its derivation are introduced for the first time in the investigation of aircraft controllability. 
  
    \item The application of these moment sets for assessing controllability in real-time is discussed.

    \item Extreme maneuvers are examined to explore the aircraft's limits, taking into account coupling and nonlinear effects. This approach contrasts with previous studies, and the effectiveness of the proposed method is demonstrated through these rigorous maneuvers.

    \item A comparison between conventional state limiters and the proposed method reveals significantly larger volumes of stable maneuverability. This result highlights the improvement in the aircraft's agility, showing enhanced maneuverability without compromising safety.
\end{itemize}

Consequently, the importance of the proposed method lies in its ability to prevent loss of control without requiring intensive offline computation and in its higher effectiveness in preserving aircraft maneuverability, with respect to the conventional state limiters. The conclusions of the study demonstrate the significant promise of the proposed method, as it enables the pilot to retain control of the aircraft in the execution of large amplitude and abrupt maneuvers.

The study is structured as follows: Section \ref{pre} establishes the necessary background, covering \ref{f16} for the modeling of the over-actuated F-16 as a baseline, and \ref{AMS} the introduction of the incremental attainable moment set for the loss of control detection. Afterwards, Section \ref{CAS} covers the design of the control augmentation system using incremental nonlinear dynamic inversion, while Section \ref{CA} discusses the optimization-based control allocation for over-actuation. The core architecture for loss of control prevention, combining Lyapunov's direct method and aircraft inverse dynamics, is detailed in Section \ref{LoC}, along with its subsystems. The effectiveness of the proposed method is demonstrated through various scenarios and aggressive maneuvers, comparing it to conventional state limiters in Section \ref{results}. Finally, Section \ref{conc} provides concluding remarks and potential future work.

\section{Preliminaries}
\label{pre}

In this section, the necessary background is established: (1) the nonlinear flight dynamics model as a baseline, and (2) the incremental attainable moment set concept, which will be leveraged for loss of control detection, as discussed in the proceeding sections.

\subsection{Nonlinear Flight Dynamics Model}
\label{f16}

The reference aircraft for the simulations is the F-16 fighter jet, equipped with an over-actuated control suite, i.e. five independent control surfaces—two ailerons (left and right), two horizontal tails (left and right), and one rudder\textemdash each capable of independent deflection directions, magnitudes, and rates based on the desired control action. Hereby, the aerodynamic modeling must be adjusted in accordance with this control suite architecture by considering the contributions of each control surface separately. Nevertheless, the over-actuated aerodynamic modeling is adopted from \cite{aerodynamicPolynomials, forceMomentEquations}. Each aerodynamic coefficient is expressed as a polynomial function involving the relevant states and control surface deflections. Also, the rigid body 6-degrees-of-freedom nonlinear equations of motion and necessary mass, inertia, actuator, and physical properties are given in \cite{F16}. As the cornerstone of this study, the nonlinear flight dynamics model is constructed based on the preceding studies in a high-fidelity manner.

\subsection{Incremental Attainable Moment Set}
\label{AMS}

The attainable moment set (AMS) of an aircraft is a geometrical representation (topologically defines a convex-hull) of the aerodynamic moments that can be generated by the aircraft's control effectors at various flight conditions. Furthermore, an attainable moment set of an aircraft helps understanding the instantaneous control authority of the aircraft, i.e. controllability of the aircraft \cite{ams1, ams2}. This direct correlation can be established under the following conditions: omitting (1) inertial couplings, (2) dynamic pressure, and (3) external factors such as atmospheric disturbances. Since these factors are not functions of the aerodynamic moment coefficients, the investigation of controllability can be distilled directly to control authority. Due to this benefit, the AMS will be used in this study to detect the possibility of loss of control. However, constructing the attainable moment set based solely on actuator position limits results in the globally achievable control moment coefficient under the specified circumstances. Instead, an alternative path is proposed to give a better insight about the instantaneous controllability during an agile maneuver, i.e. incremental attainable moment set.

The concept and utilization of the incremental attainable moment set (IAMS) is the same as the attainable moment set with a distinction: the boundary formulation. In the proposed method, both the actuator deflection limits and the actuator rate limits are considered to determine whether the aircraft can provide the necessary control moment at each time step or not. Therefore, with a slight notion shift, the attainable moment set is reformulated to construct the incremental attainable moment set $\Delta \mathbb{A_M}$ as given in Eq.~\eqref{eq:iams1}.

\begin{equation}\label{eq:iams1}
        \Delta \mathbb{A_M} = \{\Delta\tau | \Delta\tau = \Phi\Delta u, \hspace{0.15cm} \underline{\Delta u} \leq \Delta u \leq \overline{\Delta u}\}
\end{equation}
where $\Delta \tau \in \mathbb{R}^3$ is the incremental control moment coefficient matrix, whereas $\Delta u \in \mathbb{R}^m$ is the incremental control input. Furthermore, $\underline{\Delta u}$ and $\overline{\Delta u}$ denote the lower and upper boundaries of the control inputs defined as given in Eq.~\eqref{eq:IAMS1}.

\begin{equation} \label{eq:IAMS1}
    \begin{split}
    \overline{\Delta u} &= \textrm{min}(\dot{u}_{max} \Delta t_s, u_{max}-u_0) \\
    \underline{\Delta u} &= \textrm{max}(-\dot{u}_{max} \Delta t_s, u_{min}-u_0)
    \end{split}
\end{equation}
where $\dot{u}_{max}$, $u_{max}, u_{min}, u_0$ are the rate limit, maximum, minimum, and current positions of the related actuator respectively while $\Delta t_s$ is the sampling time. Eventually, using the incremental form of the control input allows to rearrange the boundaries of the problem by including the actuator rate limit; moreover, it becomes feasible to obtain more realistic attainable moment set that allows to assess the instantaneous controllability in accordance with the actuator capability.

\section{Flight Control Architecture}
\label{nfca}

This section contains the description of the design of a control augmentation system using the incremental nonlinear dynamic inversion (INDI) with the fundamental time-scale separation principle, which is a commonly preferred method \cite{indi1, indi2, indi3}. Additionally, an optimization-based control allocation design to deal with the over-actuation is discussed.

\subsection{Control Augmentation System Design}
\label{CAS}

In this study, the designed control augmentation system has three pilot commands; angle of attack ($\alpha_{cmd}$), sideslip ($\beta_{cmd}$), and body-axis roll rate ($p_{cmd}$). Such a control output scheme is regarded as more suitable for managing extreme maneuvers, as aerodynamic angles provide more information about the aircraft's aerodynamic characteristics than angular rates do. However, it is important to note that angle of attack and sideslip sensors are relatively low-rate and sensitive to atmospheric disturbances or aerodynamic interferences, which may lead to a deterioration in the performance of INDI control.

The fast dynamics and slow dynamics are decomposed in accordance with the time-scale separation. The angle of attack and sideslip are the part of the slow dynamics, whereas the body angular rates are the part of fast dynamics. Therefore, the angle of attack and sideslip are the elements of the outer-loop, and the output of the outer-loop is used as the input of the inner-loop. As a consequence, the dynamics of $\alpha$ and $\beta$ can be such decomposed as the expression given in Eq.~\eqref{eq:CAS5}.

\begin{equation} \label{eq:CAS5}
\begin{bmatrix}
\dot{\alpha} \\
\dot{\beta} \\
\end{bmatrix} 
=
\begin{bmatrix}
\dot{\alpha}_0 \\
\dot{\beta}_0 \\
\end{bmatrix}
+
G(x_0, u_0)
\begin{bmatrix}
q - q_0 \\
r - r_0 
\end{bmatrix} 
\end{equation}

If the expression is inverted to derive $q$ and $r$, the following form is obtained as given in Eq.~\eqref{eq:CAS6}.

\begin{equation} \label{eq:CAS6}
\begin{bmatrix}
q \\
r 
\end{bmatrix} 
=
G(x_0, u_0)^{-1}\Bigg(
\begin{bmatrix}
\dot{\alpha} \\
\dot{\beta} \\
\end{bmatrix} 
-
\begin{bmatrix}
\dot{\alpha}_0 \\
\dot{\beta}_0 \\
\end{bmatrix} 
\Bigg)
+
\begin{bmatrix}
q_0 \\
r_0 
\end{bmatrix} 
\end{equation}
which $G(x_0, u_0)$ is given in Eq.~\eqref{eq:CAS7}. 

\begin{equation} \label{eq:CAS7}
G(x_0, u_0) =
\begin{bmatrix}
1-\frac{1}{4m\cos\beta}\rho S \Bar{c} C_{L_{q}} & -\tan\beta \sin\alpha \\
\frac{\rho S \Bar{c} C_{D_{q}}}{4m} \sin\beta (1-\cos\beta) & \frac{\rho S b C_{Y_{r}}}{4m} \cos^2 \beta - \cos\alpha \\
\end{bmatrix} 
\end{equation}
where $\rho, \Bar{c}, b, S$ are the density, the mean aerodynamic chord length, the wing span, and the wing area respectively. Also, $C_{L_{q}}, C_{D_{q}}, C_{Y_{r}}$ are the lift and drag coefficient derivatives with respect to the non-dimensional pitch rate, and side force coefficient derivative with respect to the non-dimensional yaw rate, respectively. Moreover, $[\dot{\alpha} \hspace{0.15cm} \dot{\beta}]^{T}$ is the virtual input and can be obtained using a simple proportional controller as given in Eq.~\eqref{eq:CAS8}.

\begin{equation} \label{eq:CAS8}
\begin{bmatrix}
\dot{\alpha} \\
\dot{\beta} \\
\end{bmatrix}
= 
\begin{bmatrix}
\omega_\alpha(\alpha_{cmd} - \alpha) \\
\omega_\beta(\beta_{cmd} - \beta) \\
\end{bmatrix} 
\end{equation}
where $\omega_\alpha$ and $\omega_\beta$ are the proportional gains for $\alpha$ and $\beta$ channels, respectively. Eventually, the pilot commands for angle of attack ($\alpha_{cmd}$) and sideslip angle ($\beta_{cmd}$) are converted into desired angular rates, specifically roll rate ($p$), pitch rate ($q$), and yaw rate ($r$), which serve as inputs for the inner-loop control system, which will be illustrated in Fig~\ref{fig:controlAllocationDiagram}.

Subsequent to finalizing the outer-loop, the inner-loop is designed based on angular rates; $p$, $q$, and $r$. Furthermore, the outputs of the inner-loop are the control moment coefficients rather than the control surface deflections due to the control allocation. The required moment coefficients can be derived using Euler's equations of motion as given in Eq.~\eqref{eq:CAS9}.

\begin{equation} \label{eq:CAS9}
\begin{split}
\begin{bmatrix}
C_l \\
C_m \\
C_n
\end{bmatrix}
=
\frac{J}{\frac{1}{2}\rho V^2 S}
\begin{bmatrix}
b & 0 & 0 \\
0 & \Bar{c} & 0 \\
0 & 0 & b
\end{bmatrix}^{-1}
\Bigg\{
\begin{bmatrix}
\dot{p} \\
\dot{q} \\
\dot{r} 
\end{bmatrix}
 + J^{-1}
\Bigg(
\begin{bmatrix}
p \\
q \\
r 
\end{bmatrix}
\times
J
\begin{bmatrix}
p \\
q \\
r 
\end{bmatrix}
\Bigg)
\Bigg\}
\end{split}
\end{equation}
which ${[C_l \hspace{0.15cm} C_m \hspace{0.15cm} C_n]^{T}}$ corresponds to the roll, pitch, and yaw moment coefficients as the outputs of the inner-loop. In addition, ${[\dot{p} \hspace{0.15cm} \dot{q} \hspace{0.15cm} \dot{r}]^{T}}$ is the virtual input. As derived previously, the virtual input is given in Eq.~\eqref{eq:fca13}.

\begin{equation} \label{eq:fca13}
\begin{bmatrix}
\dot{p} \\
\dot{q} \\
\dot{r}
\end{bmatrix}
= 
\begin{bmatrix}
\omega_p(p_{cmd} - p) \\
\omega_q(q_{cmd} - q) \\
\omega_r(r_{cmd} - r) \\
\end{bmatrix} 
\end{equation}
where $\omega_p, \omega_q, \omega_r$ are the proportional gains for $p, q$ and $r$ channels, respectively. Note that $q_{cmd}$ and $r_{cmd}$ are derived from the commanded angle of attack and sideslip in the outer loop, whereas $p_{cmd}$ is the direct pilot commands. The proportional gains corresponding to each control channel in the proportional controllers are given in Table~\ref{tb:gainsLC}.

\begin{table}[hbt!]
\caption{Proportional controller gains for each control channel}
\centering
\begin{tabular}{ccccc}
\hline
\hline
$\omega_\alpha$ & $\omega_\beta$ & $\omega_p$ & $\omega_q$ & $\omega_r$ \\
\hline
2.5 & 1 & 10 & 10 & 5 \\
\hline
\hline
\end{tabular}
\label{tb:gainsLC}
\end{table}

\subsection{Control Allocation Design}
\label{CA}

The need for a control allocation algorithm due to over-actuation has already been discussed. Thus, to calculate the necessary control surface deflections, an optimization-based control allocation is utilized. The objective function represents the minimization of the demand on the control system, with
control surfaces position limiting. Additionally, the demand on the control system is quantified using the square of the Euclidean 2-norm, denoted as $||u||^2_{2}$ or $u^T u$ in the vector form. Specifically, the flight control allocation problem can be formulated as given in Eq.~\eqref{eq:CA1}.

\begin{equation} \label{eq:CA1}
    \begin{aligned}
    \min_{u} \quad & \frac{1}{2}u^T u \\
    \textrm{s.t.} \quad & \norm{\Phi(x_0, u_0) u - \tau_c} = 0 \\
                  \quad & u_{min} \leq  u \leq u_{max}
    \end{aligned}
\end{equation}
where $\Phi(x_0, u_0) \in \mathbb{R}^{3 \times m}$ is the Jacobian matrix at that instant, where $m$ is the number of control surfaces, and $\tau_c$ represents the commanded control moment coefficients. Additionally, the upper and lower bounds of the actuator position limits, represented by $u_{min}$ and $u_{max}$, significantly influence the derived control surface deflections. Furthermore, the Jacobian matrix is computed at each time step using the central-difference method. This approach captures the aerodynamic nonlinearity through discretization, accommodating the nonlinear and complex aerodynamic characteristics. 

\begin{figure}[hbt!]
 \centering
 \includegraphics[width=6.5in]{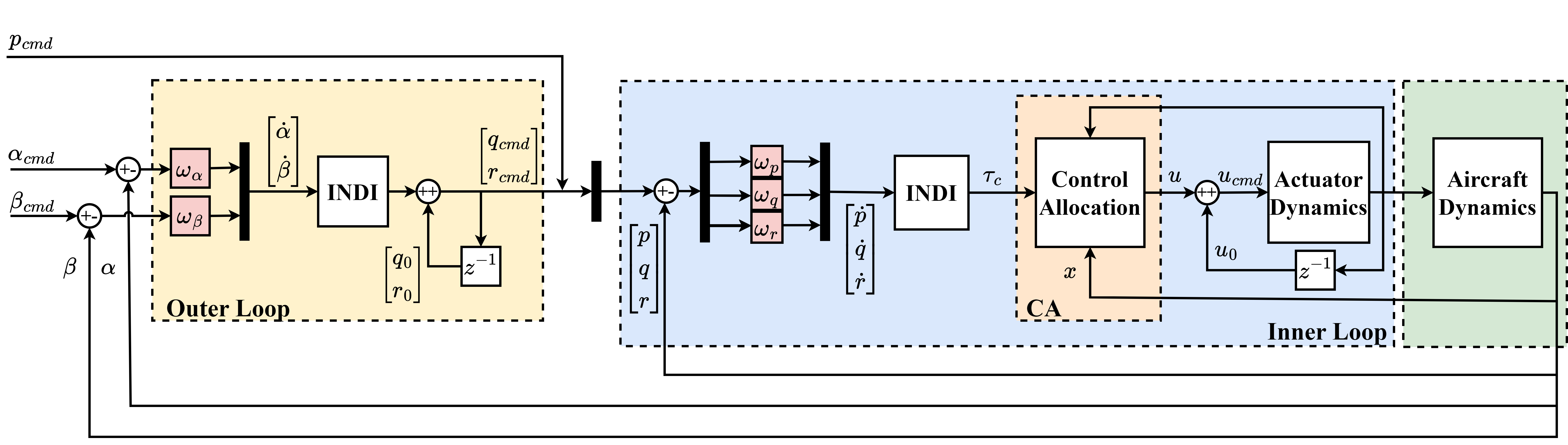}
 \caption{Entire flight control architecture.} 
 \label{fig:controlAllocationDiagram}
\end{figure}

The desired control moment coefficient can be achieved by minimizing the control effort with the help of the presented formulation. however, another concern regarding optimization-based control allocation is the selection of a stable, fast-converging, and reliable optimization algorithm. In this study, the  selected optimization algorithm is sequential-quadratic programming due to its ability to solve highly complex optimization problems with high reliability, in addition to its stable convergence attitude \cite{yca9, yca11}. Subsequent to finalizing the discussion of the control allocation design, for the sake of clarity, a flowchart representing the entire flight control architecture is illustrated in Fig.~\ref{fig:controlAllocationDiagram}.

As illustrated in Fig.~\ref{fig:controlAllocationDiagram}, the pilot commands $\alpha_{cmd}$, $\beta_{cmd}$, and $p_{cmd}$ are subjected to the incremental nonlinear dynamic inversion to derive the control moment coefficients. Afterwards, the control moment coefficient are transferred to the control allocation module to derive the corresponding control surface deflections while minimizing the control effort. All in all, the control surface deflections are injected into the nonlinear flight dynamics module for deriving new states, $x$. Notice that, since the inner-loop is also derived using incremental nonlinear dynamic inversion, the finalized control surface deflections are also in the incremental form. This implies that the control allocation is only responsible for distributing the necessary control moment coefficients on the control surfaces in an optimal manner, as the main objective is minimizing the control effort while satisfying the desired control moment coefficients previously described.

\section{Loss of Control Prevention}
\label{LoC}

The loss of control notion has been discussed in the context of controllability, and this study specifically addresses the prevention of departure. Furthermore, the architecture is divided into two stages: detection and prevention, both of which are proposed in an online manner.

\begin{itemize}
    \item \textbf{Detection:} The IAMS serves as the detector for the initiation of the loss of control, indicating the available control authority. The flight control architecture derives the desired control moment vector, which is then compared to the instantaneous IAMS to determine whether the demand falls within the bounds of the available control authority. However, as a safety margin, the acquired instantaneous IAMS is shrunk by a factor of 30\%, meaning that only 70\% of the available control authority is allowed to be used. This direct correlation between the IAMS and the available control authority is achievable, provided that the aerodynamic database is extensive and high-fidelity, accounting for nonlinearities and over-actuation effects as comprehensively as possible. Moreover, the utilized shrinkage factor is not a strict requirement but can be modified based on expectations (the greater the shrinkage factor, the more restrictive it is). The specified level is regarded as sufficiently satisfactory in this study, considering the trade-off between maneuverability and loss of control. Hence, the approach will be conservative when a moment cannot be instantaneously achieved, and the prevention mechanism in the controller saturates the command.
    \item \textbf{Prevention:} An online command saturation approach is developed in order to prevent the loss of control (i.e. departure) due to the excessive or abrupt pilot inputs. In case of the detection of the violation of the control authority, the online command saturation module is activated and the states of $\alpha$, $p$, $q$ and $r$ are restricted. The online command saturation is designed based on the Lyapunov's direct method, which will be discussed in the proceeding section.
\end{itemize}

Consequently, the primary objective is to prevent departure during extreme maneuvers by saturating the pilot commands in the event of a loss of control detection. In this way, executing maneuvers as agilely as the proposed strategy permits without entering into an upset condition is desired.

\subsection{Lyapunov-based Online Command Saturation Design}
\label{dynSaturation}

The Lyapunov stability theorem has been widely used in the control system design to ensure that a system remains stable in the face of disturbances or uncertainties, and to design control strategies that can stabilize an unstable system. In this section, the derivation of the command saturation will be discussed. Principally, set such a Lyapunov candidate as given in Eq.~\eqref{eq:dynSaturation1}.

\begin{equation}
\label{eq:dynSaturation1}  
    V(\omega) = \omega^T J \omega
\end{equation}
where $\omega = [p \hspace{0.15cm} q \hspace{0.15cm} r]^T$ is the angular velocity vector and $J$ is the inertia tensor. Actually, the selected Lyapunov function is an energy-like function, i.e. rotational kinetic energy, and $V(\omega) > 0, \forall \omega \in \mathbb{R}^n$ since $J \in \mathbb{R}^{3 \times 3}$ is a symmetric positive-definite matrix. Moreover, if $\omega^* = [0 \hspace{0.15cm} 0 \hspace{0.15cm} 0]^T$ is the equilibrium point, $V(\omega^*) = 0$. If the time derivative of the Lyapunov candidate is taken as given in Eq.~\eqref{eq:dynSaturation2},

\begin{equation}
\label{eq:dynSaturation2}  
    \Dot{V}(\omega) = 2\omega^T J \Dot{\omega}
\end{equation}
can be obtained. Recall the Euler's equations of motion in a compact form as given in Eq.~\eqref{eq:dynSaturation3}.

\begin{equation}\label{eq:dynSaturation3}
    J\Dot{\omega} = -[\omega \times J\omega] + M    
\end{equation}

If Eq.~\eqref{eq:dynSaturation3} is plugged in Eq.~\eqref{eq:dynSaturation2}, the following form is obtained as given in Eq.~\eqref{eq:dynSaturation4}.

\begin{equation}\label{eq:dynSaturation4}
    \Dot{V}(\omega) = 2\omega^T J \Dot{\omega} = 2\omega^T (-[\omega \times J \omega] + M)    
\end{equation}

The term $2\omega^T M$ must be negative, as the criterion $\Dot{V}(\omega) < 0$ is required for asymptotic stability. Thus, $\dot{V}$ can be transformed to $2\omega^T M$, and a moment design specified in Eq.~\eqref{eq:dynSaturation5} is needed to achieve asymptotic stability in the system.

\begin{equation}
\label{eq:dynSaturation5}  
    M = -K \omega
\end{equation}
where $K \in \mathbb{R}^{3 \times 3}$ is symmetric positive-definite weight matrix. If Eq.~\eqref{eq:dynSaturation5} is plugged in Eq.~\eqref{eq:dynSaturation4}, the following is obtained. 

\begin{equation}
\label{eq:dynSaturation6}  
    \Dot{V}(\omega) = -2\omega^T K \omega < 0
\end{equation}

Consequently, the stabilizing control moments in Eq.~\eqref{eq:dynSaturation5} are derived using Lyapunov's direct method, and these regulator moments will be utilized to derive the pilot command saturation expressions. By using Euler's equations of motion, the angular rate saturation can be established through the following relation given in Eq.~\eqref{eq:dynSaturation7}.

\begin{equation} \label{eq:dynSaturation7}
\underbrace{\begin{bmatrix}
L \\
M \\
N
\end{bmatrix}}_{\substack{-K
\begin{bmatrix}
p \\
q \\
r
\end{bmatrix}
}}
=
J
\underbrace{\begin{bmatrix}
\dot{p} \\
\dot{q} \\
\dot{r} 
\end{bmatrix}_{sat}}_{\substack{
\begin{bmatrix}
\omega_p (p_{sat} - p) \\
\omega_q (q_{sat} - q) \\
\omega_r (r_{sat} - r)
\end{bmatrix}
}}
+
\begin{bmatrix}
p \\
q \\
r 
\end{bmatrix}
\times
J
\begin{bmatrix}
p \\
q \\
r 
\end{bmatrix}
\end{equation}

Obviously, the stabilizing moments can be converted into the angular rates as given in the Eq.~\eqref{eq:dynSaturation8}. 

\begin{equation} \label{eq:dynSaturation8}
J^{-1}\Bigg\{-K
\begin{bmatrix}
p \\
q \\
r
\end{bmatrix}
-
\begin{bmatrix}
p \\
q \\
r 
\end{bmatrix}
\times
J
\begin{bmatrix}
p \\
q \\
r 
\end{bmatrix}
\Bigg\}
=
\begin{bmatrix}
\omega_p (p_{sat} - p) \\
\omega_q (q_{sat} - q) \\
\omega_r (r_{sat} - r)
\end{bmatrix}
\end{equation}

Eventually, if $[p \hspace{0.15cm} q \hspace{0.15cm} r]^T_{sat}$ is decomposed from the expression given in Eq.~\eqref{eq:dynSaturation8}, the following Eq.~\eqref{eq:dynSaturation9} can be derived.

\begin{equation} \label{eq:dynSaturation9}
\begin{bmatrix}
p_{sat} \\
q_{sat} \\
r_{sat}
\end{bmatrix}
=
\Bigg\{
J^{-1}\Bigg\{-K
\begin{bmatrix}
p \\
q \\
r
\end{bmatrix}
-
\begin{bmatrix}
p \\
q \\
r 
\end{bmatrix}
\times
J
\begin{bmatrix}
p \\
q \\
r 
\end{bmatrix}
\Bigg\}
\Bigg\}
\oslash
\begin{bmatrix}
\omega_p \\
\omega_q \\
\omega_r 
\end{bmatrix}
+
\begin{bmatrix}
p \\
q \\
r
\end{bmatrix}
\end{equation}
where $\oslash$ denotes the Hadamard division. The obtained $[p \hspace{0.15cm} q \hspace{0.15cm} r]^T_{sat}$ is derived from the stabilizing moments, indicating that $[p \hspace{0.15cm} q \hspace{0.15cm} r]^T_{sat}$ is also stabilizing. In other words, this represents the set of states corresponding to the limits of the controllable envelope, which depends on the designed Lyapunov gain $K$ given by Eq.~\eqref{eq:dynSaturation5}. Furthermore, to design the outer-loop saturation, remember the following relation given in Eq.~\eqref{eq:dynSaturation10} which is derived using the incremental nonlinear dynamic inversion.

\begin{equation} \label{eq:dynSaturation10}
\underbrace{\begin{bmatrix}
\dot{\alpha} \\
\dot{\beta} \\
\end{bmatrix}}_{\substack{
\begin{bmatrix}
\omega_\alpha (\alpha_{sat} - \alpha) \\
\omega_\beta (\beta_{sat} - \beta) \\
\end{bmatrix}
}}
=
\begin{bmatrix}
\dot{\alpha}_0 \\
\dot{\beta}_0 \\
\end{bmatrix}
+
G(x_0, u_0)
\begin{bmatrix}
q - q_0 \\
r - r_0 
\end{bmatrix} 
\end{equation}

As a result, if $[\alpha \hspace{0.15cm} \beta]^T_{sat}$ is decomposed from the expression given in Eq.~\eqref{eq:dynSaturation10}, the following Eq.~\eqref{eq:dynSaturation11} is derived.

\begin{equation} \label{eq:dynSaturation11}
\begin{bmatrix}
\alpha_{sat} \\
\beta_{sat} \\
\end{bmatrix}
=
\Bigg\{
\begin{bmatrix}
\dot{\alpha}_0 \\
\dot{\beta}_0 \\
\end{bmatrix}
+
G(x_0, u_0)
\begin{bmatrix}
q - q_0 \\
r - r_0 
\end{bmatrix} 
\Bigg\}
\oslash
\begin{bmatrix}
\omega_\alpha \\
\omega_\beta \\
\end{bmatrix}
+
\begin{bmatrix}
\alpha \\
\beta \\
\end{bmatrix}
\end{equation}

\begin{figure*}[hbt!]
\centering
\includegraphics[width=\textwidth]{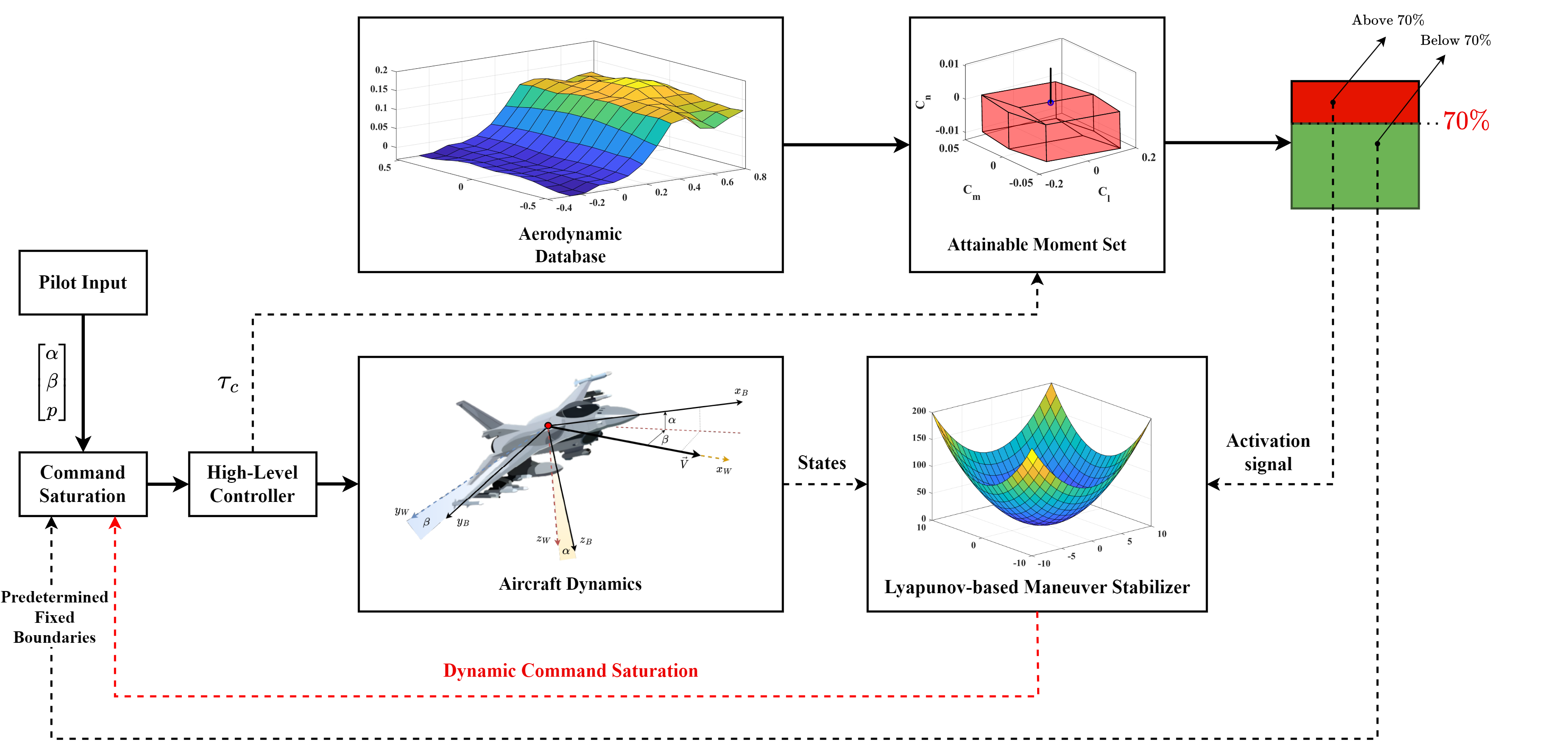}
\caption{Overview of the proposed loss of control prevention logic}
\label{fig:overviewLOClogic}
\end{figure*}

In order to make $[\alpha \hspace{0.15cm} \beta]^T_{sat}$ stabilizing, the expression of $[q \hspace{0.15cm} r]^T$ should be converted into the expression of $[q \hspace{0.15cm} r]_{sat}^T$.

\begin{equation} \label{eq:dynSaturation12}
\begin{bmatrix}
\alpha_{sat} \\
\beta_{sat} \\
\end{bmatrix}
=
\Bigg\{
\begin{bmatrix}
\dot{\alpha}_0 \\
\dot{\beta}_0 \\
\end{bmatrix}
+
G(x_0, u_0)
\begin{bmatrix}
q_{sat} - q_0 \\
r_{sat} - r_0 
\end{bmatrix} 
\Bigg\}
\oslash
\begin{bmatrix}
\omega_\alpha \\
\omega_\beta \\
\end{bmatrix}
+
\begin{bmatrix}
\alpha \\
\beta \\
\end{bmatrix}
\end{equation}
where the expression of $[q \hspace{0.15cm} r]_{sat}^T$ is derived in Eq.\eqref{eq:dynSaturation9}. However, only the angle of attack command is intervened since the sideslip angle is commanded to be zero in all scenarios. This assumption is based on the expectation that a zero sideslip command can be tracked with only minor deviations. Furthermore, since Eq.~\eqref{eq:dynSaturation12} accounts for instantaneous states, the effects of a non-zero sideslip can be reflected to the yielded stabilizing angle of attack command value. Consequently, through inductive reasoning, the expression for online command saturation can be derived. By designing the stabilizing control moments, a correlation between them and the online command saturation expressions can be established. Therefore, an online solution for the command saturation is proposed. Implementing the proposed online architecture requires accurate real-time measurements and a high-fidelity aerodynamic database that covers a broad validity envelope, including post-stall conditions, with fine discretization and sufficient margins to account for the full range of the aircraft’s service envelope. The remaining steps involve simple algebraic calculations, so the computational cost depends only on deriving the necessary aerodynamic coefficients. After all, for the sake of clarity, a flowchart of the proposed logic is presented in Fig.~\ref{fig:overviewLOClogic}.

\section{Results}
\label{results}

The proposed method is reviewed under different tasks and circumstances. In Section~\ref{abruptAgileMan}, the ability of the loss of control prevention is analyzed under harsh maneuver tasks and alternating initial conditions. In Section~\ref{comparisonLyapunov}, moreover, the abilities of loss of control prevention and sustaining a stable flight are scrutinized by comparing the proposed method to a conventional scheduled state limiter approach. Note that, the simulations are conducted using a \nth{4}-order Runge-Kutta integration scheme in the MATLAB,\textsuperscript{\tiny\textregistered} Simulink environment.

\subsection{Abrupt \& Agile Maneuver Evaluation}
\label{abruptAgileMan}

The preliminary assessments of the proposed architecture are presented with two distinct agile maneuvers\footnote{For the simulation videos, please visit: https://youtu.be/I524wsm6EPk}. The agility of the aircraft is evaluated based on the following command metrics: (1) high amplitude control commands (commands are considered high amplitude if they involve changes greater than 30$^\circ$ for pitch rate and 90$^\circ$ for roll rate); (2) abrupt application (commands are considered abrupt if they change considerably in a manner similar to step-wise changes); (3) simultaneous application (commands are considered simultaneous if they are applied concurrently). Based on these metrics, angle of attack and roll rate inputs are applied abruptly, with high amplitude, and simultaneously. Furthermore, the static stability degradation has a significant impact on the loss of control; that is, the worst-case scenario is selected, and the center of gravity is located such that the aircraft is statically unstable ($x_{CG}$ = $\%35\Bar{c}$). Subsequently, the effects of the proposed architecture are presented after demonstrating that the aircraft is not capable of performing the desired maneuver without the loss of control prevention architecture. The flight conditions and the $x_{CG}$ position set for the simulations are detailed in Table~\ref{tb:concManeuvers}.

\begin{table}[hbt!]
\begin{center}
\caption{Maneuver properties}\label{tb:concManeuvers}
\begin{tabular}{cccc}
\hline
\hline
Maneuver ID & \multicolumn{2}{c}{Initial conditions} & CG, $\%$ \\ 
\cline{2-3}
  & Mach & Altitude, m &  \\
\hline
1 & 0.8 & 2000 & $\%35\Bar{c}$ \\
2 & 1 & 500 & $\%35\Bar{c}$ \\
\hline
\hline
\end{tabular}
\end{center}
\end{table}

The proposed method is examined under various initial conditions, i.e. different dynamic pressure and control authority levels in order to understand the sensitivity of the method's performance to the alteration of the initial conditions.

\subsubsection{Maneuver \#1 Examination}

\begin{figure}[hbt!]
\centering
\includegraphics[width=5.3in]{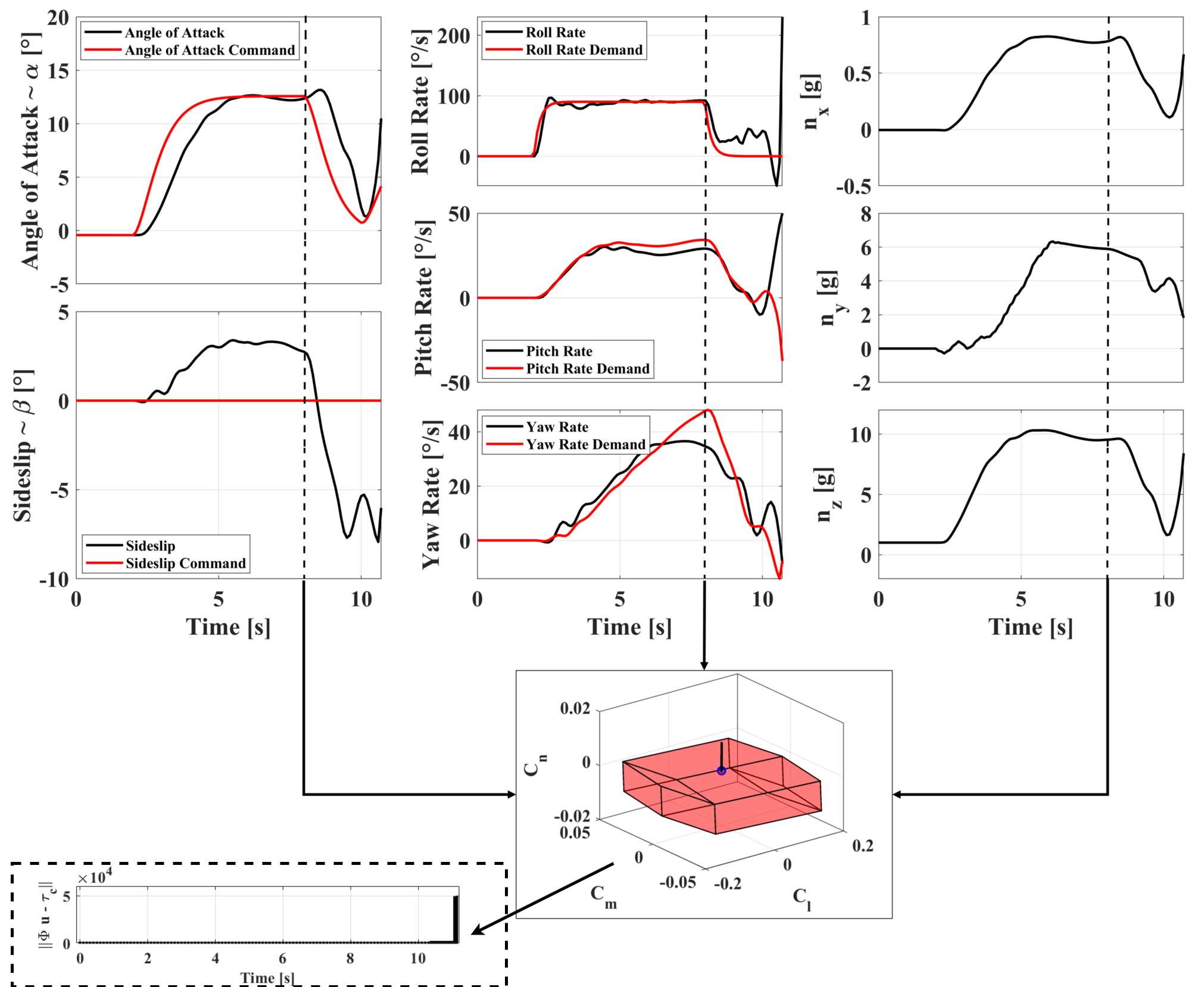}
\caption{State trajectory of the maneuver \#1 without loss of control prevention.}
\label{fig:statetrajectoryMan1woutLOCprevention}
\end{figure}

Initially, the maneuver performance of the aircraft without the loss of control prevention architecture is depicted in Fig.~\ref{fig:statetrajectoryMan1woutLOCprevention}.

%%%% without LOC prevention figures %%%%

\begin{figure}[hbt!]
\centering
\includegraphics[width=6.2in]{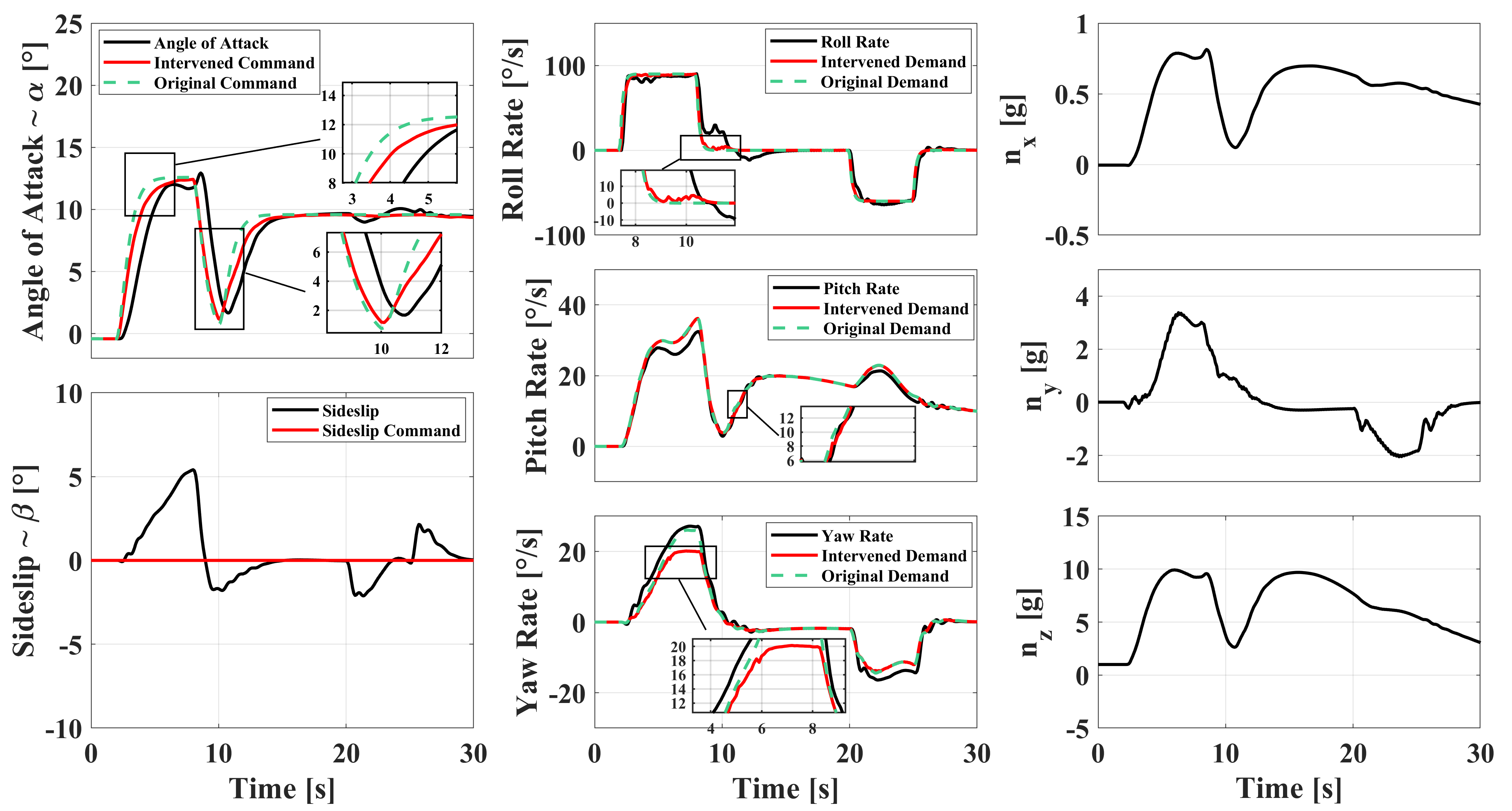}
\caption{State trajectory of the maneuver \#1 with loss of control prevention: indicates the intervention of the Lyapunov assessment.}
\label{fig:statetrajectoryMan1}
\end{figure}

Obviously, the aircraft is incapable of executing the demanded maneuver due to the violation of the controllable space, i.e., the attainable moment set, at the instant of input reversals. The violation of controllability boundaries initiates a departure from stable flight, resulting in the occurrence of loss of control. Subsequent to stimulation of the loss of control, the control deflections grow irrationally and even reach saturation. Additionally, the discrepancy between the required and achieved control moment coefficients progressively worsens until the simulation is terminated due to numerical issues arises from the loss of control. However, the effects of the loss of control prevention architecture implementation are shown in Fig.~\ref{fig:statetrajectoryMan1}.

%%%% with LOC prevention figures %%%%

\begin{figure}[hbt!]
\centering
\includegraphics[width=6in]{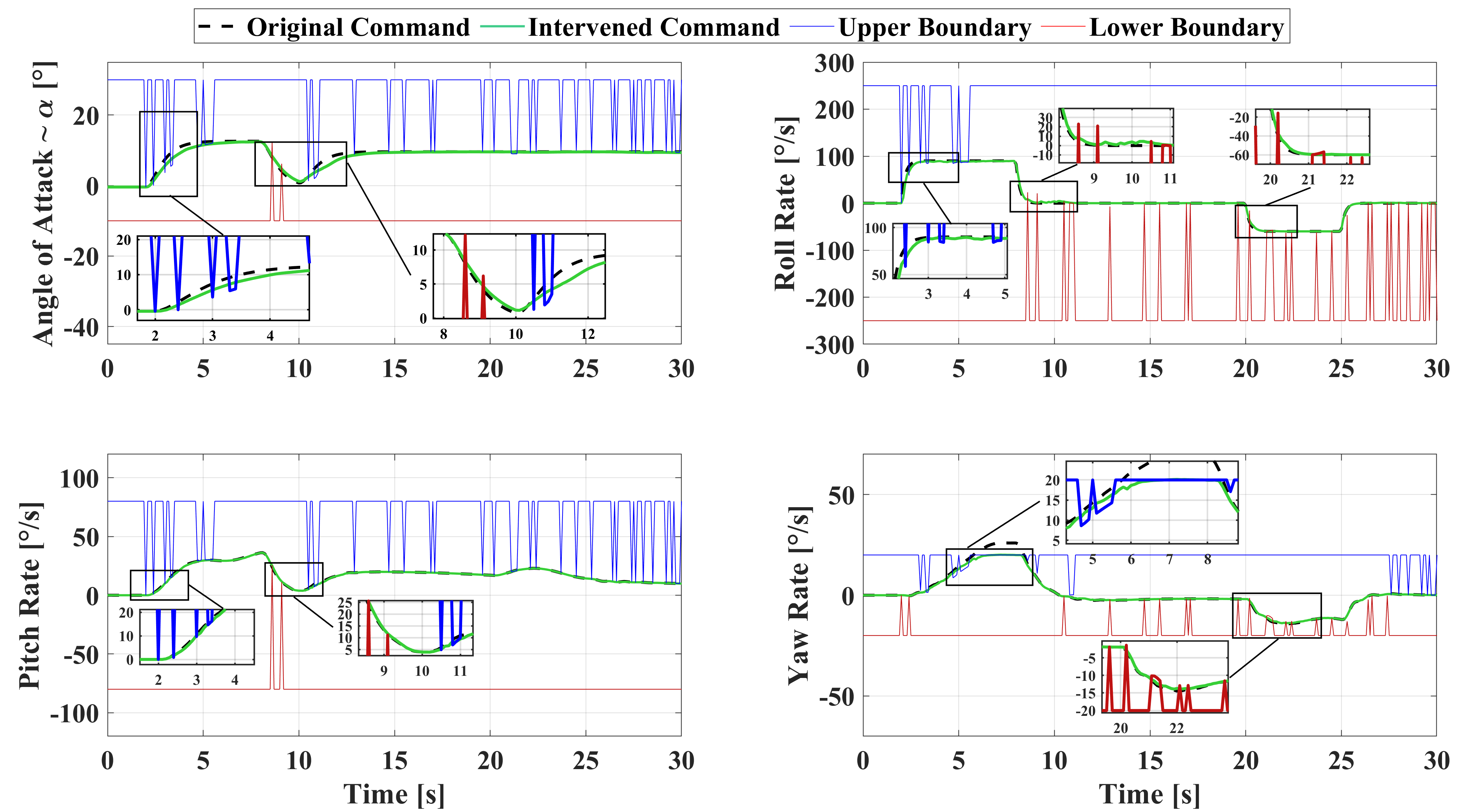}
\caption{Command saturation trajectory of the maneuver \#1 with loss of control prevention: indicates the intervention of the Lyapunov assessment.}
\label{fig:boundarytrajectoryMan1}
\end{figure}

Furthermore, the variations in the saturation of the corresponding states of the aircraft are demonstrated in Fig.~\ref{fig:boundarytrajectoryMan1}.

\begin{figure}[hbt!]
  \centering
  \begin{tabular}[b]{c}
    \includegraphics[width=3.0in]{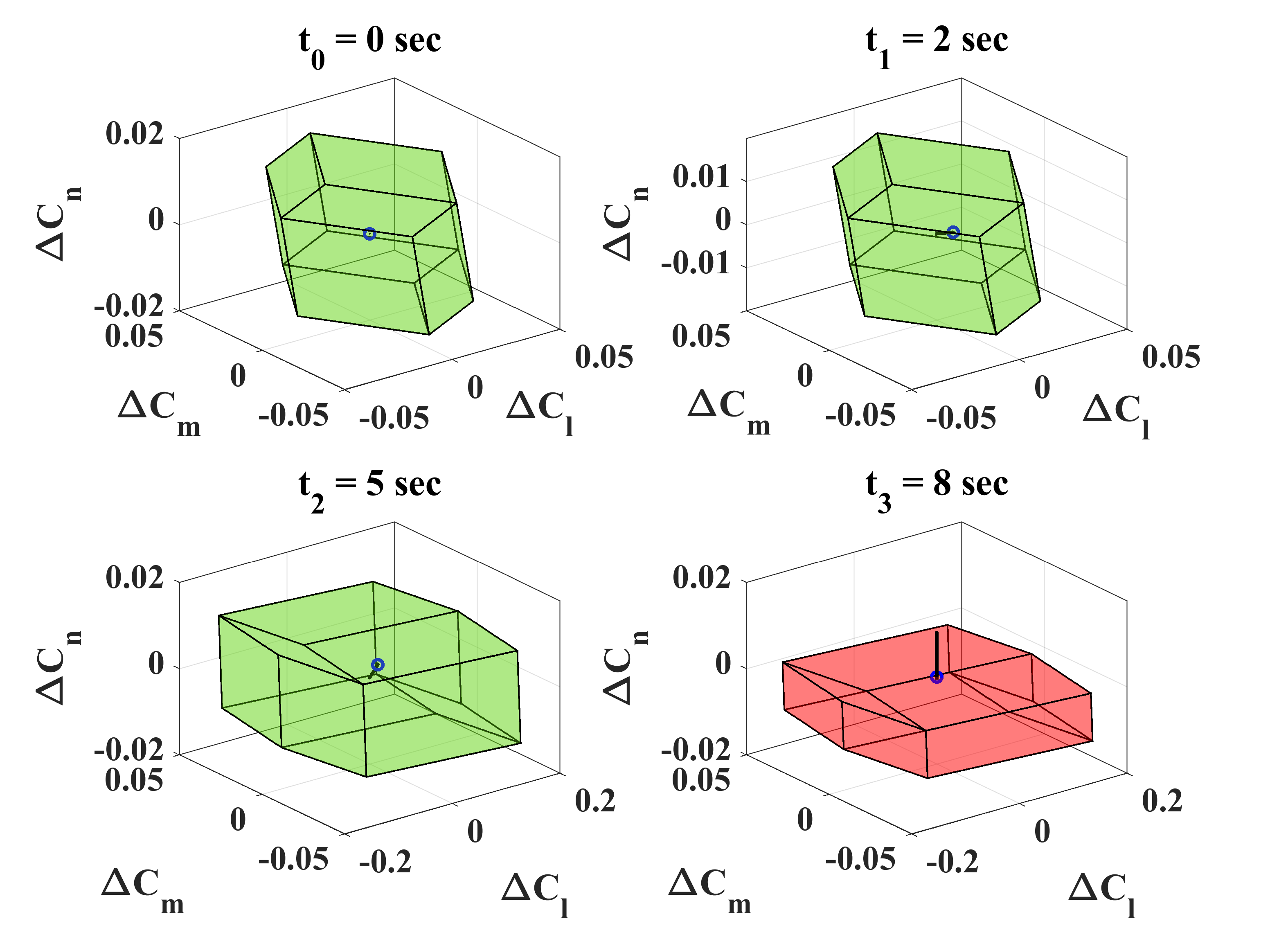} \\
    \small (a) Without loss of control prevention 
  \end{tabular} 
  \begin{tabular}[b]{c}
    \includegraphics[width=3.0in]{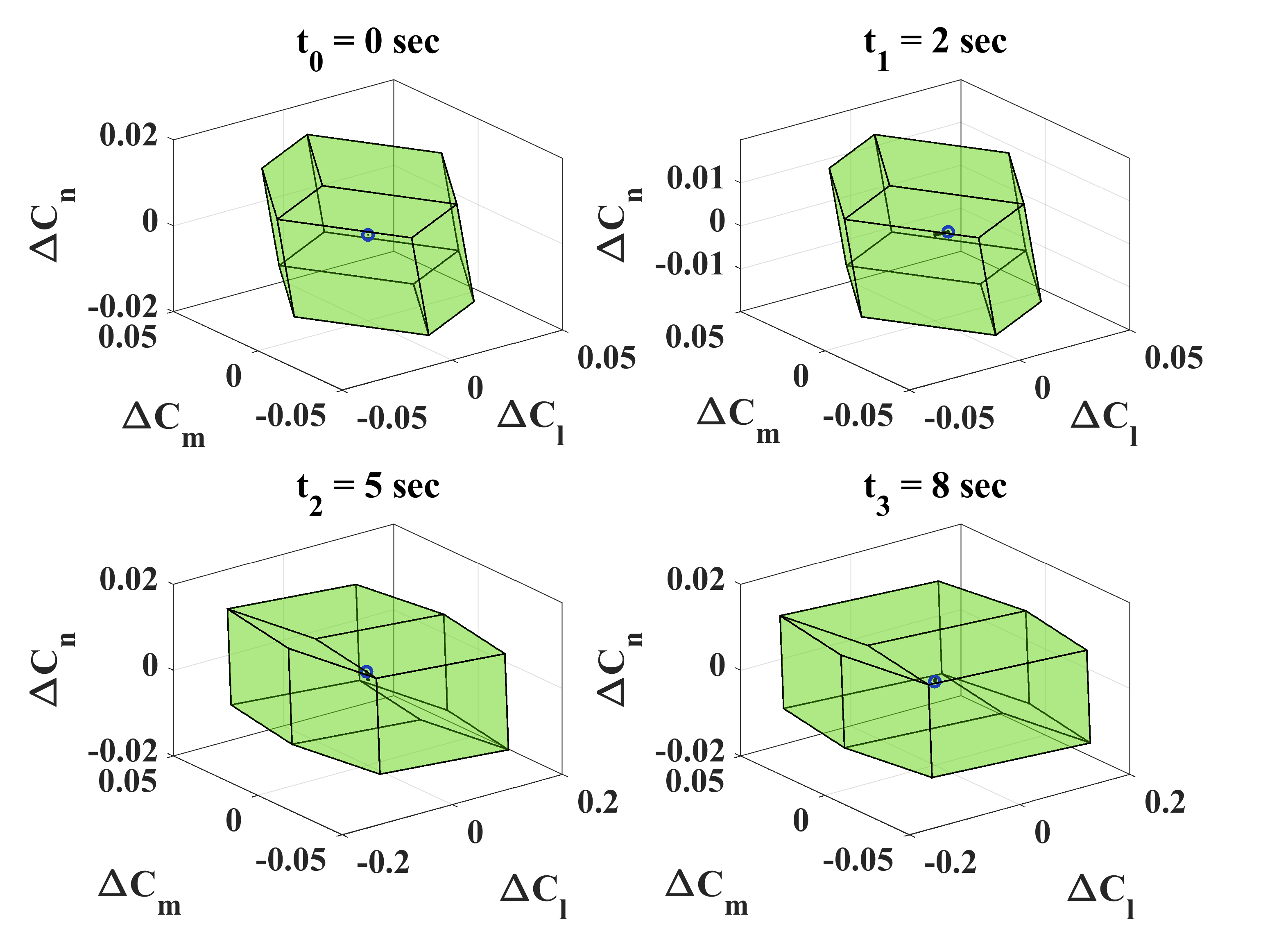} \\
    \small (b) With loss of control prevention
  \end{tabular} 
  \caption{Attainable moment set history: (a) without loss of control prevention, (b) with loss of control prevention during maneuver \#1.}
  \label{fig:man1_5}
\end{figure}

The history of the attainable moment set with loss of control prevention both enabled and disabled is shown in Fig.~\ref{fig:man1_5}. 

\begin{figure}[hbt!]
 \centering
 \includegraphics[width=5in]{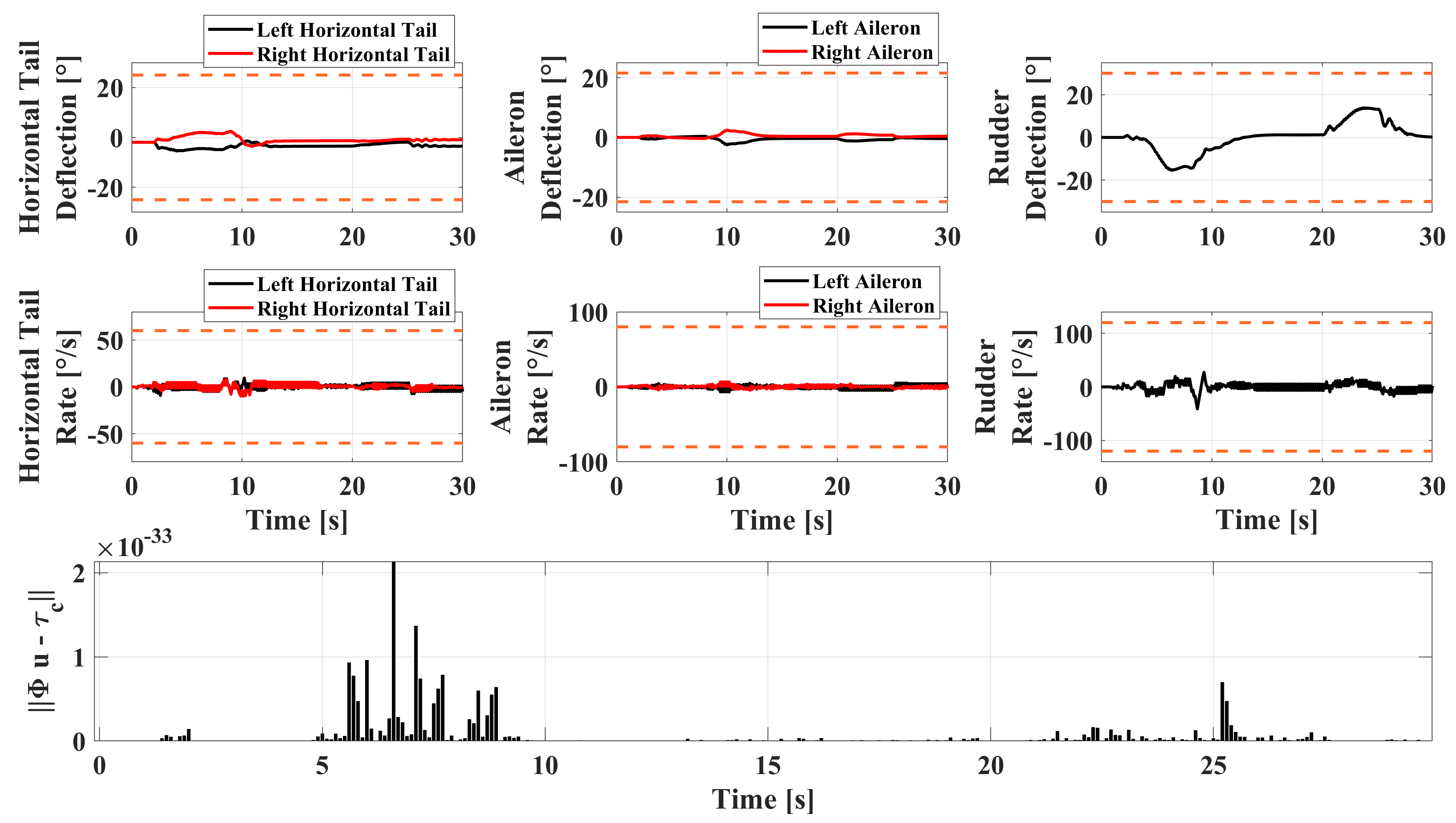}
 \caption{Control deflections with demanded moment satisfaction parameter of the maneuver \#1 with loss of control prevention.}
 \label{fig:controlMan1}
\end{figure}

\begin{figure}[hbt!]
 \centering
 \includegraphics[width=3.75in]{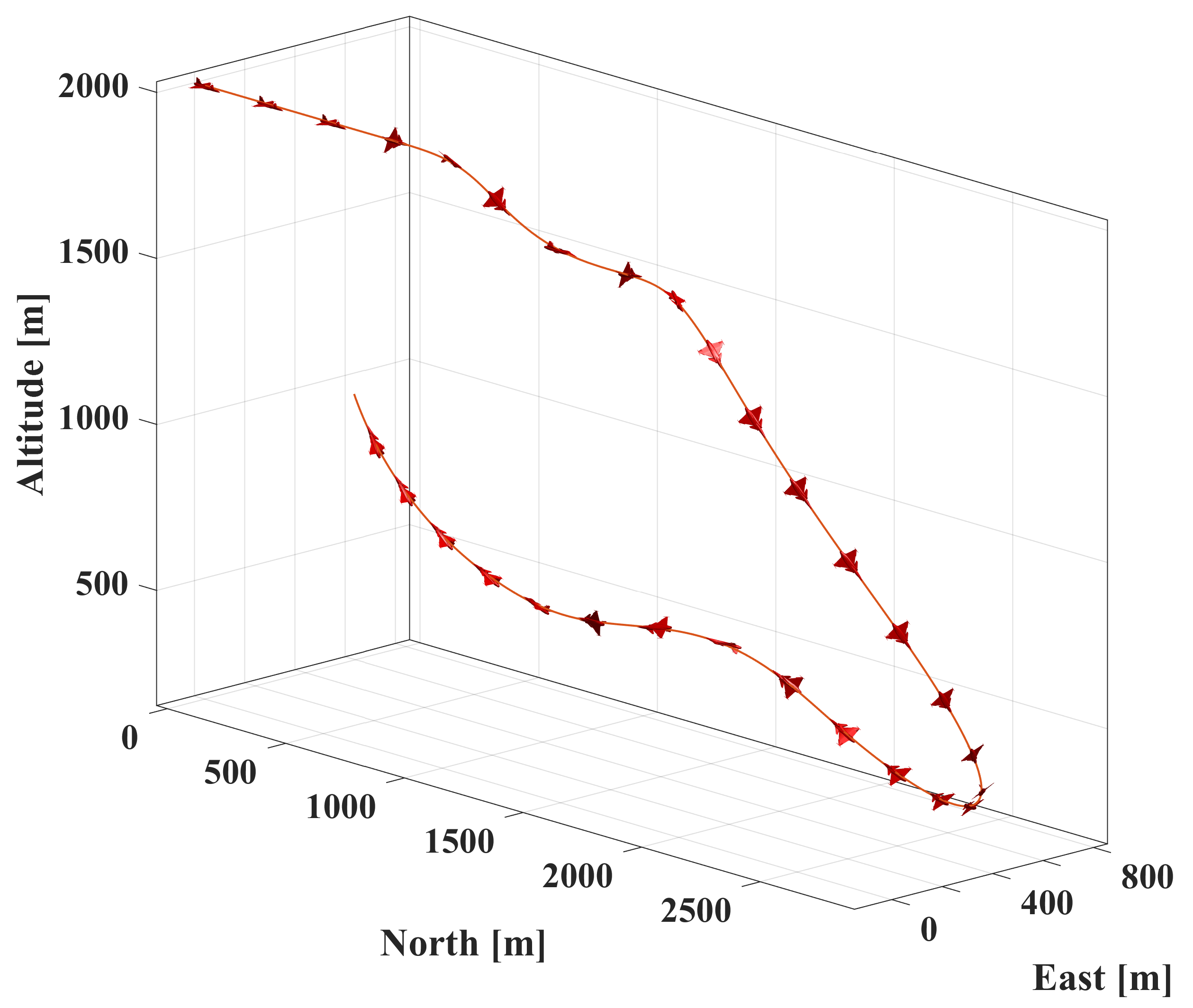}
 \caption{3D trajectory of the maneuver \#1 with loss of control prevention.}
 \label{fig:3DtrajMan1}
\end{figure}

Attainable moment set illustrations of the maneuver \#1 with loss of control prevention indicates the arbitrary instantaneous controllability checks as stated previously, and it is observed that at $t_3$ the desired control moment vector violates the attainable moment set, i.e. red, while there is not loss of control prevention. Remember from Fig.~\ref{fig:statetrajectoryMan1woutLOCprevention} that the violation stimulates the loss of control and the aircraft departures. However, in case of activation of the loss of control prevention architecture, the demanded control moment coefficient vector is still bounded by the attainable moment set; therefore, any loss of control is avoided. Consequently, the history of the control deflections with demanded moment satisfaction parameter and the 3D trajectory of the maneuver are depicted in Fig.~\ref{fig:controlMan1} and Fig.~\ref{fig:3DtrajMan1} respectively.

The control surface deflection history in Fig.\ref{fig:controlMan1} puts forth that the desired control moment coefficients are achieved with remarkable accuracy, while the magnitudes of the control surface deflections remain relatively modest, indicating that the control effort is minimized as much as possible. Obviously, without the loss of control prevention architecture, the aircraft is unable to execute the desired maneuver effectively. The simulation results demonstrate that with the loss of control prevention enabled, the aircraft successfully performs the maneuver, in contrast to the unsuccessful attempt without this feature. It is important to note that the aircraft exhibits such high agility during the maneuver that the load factor approaches the 10g limit.

\subsubsection{Maneuver \#2 Examination}

\begin{figure}[hbt!]
\centering
\includegraphics[width=5.3in]{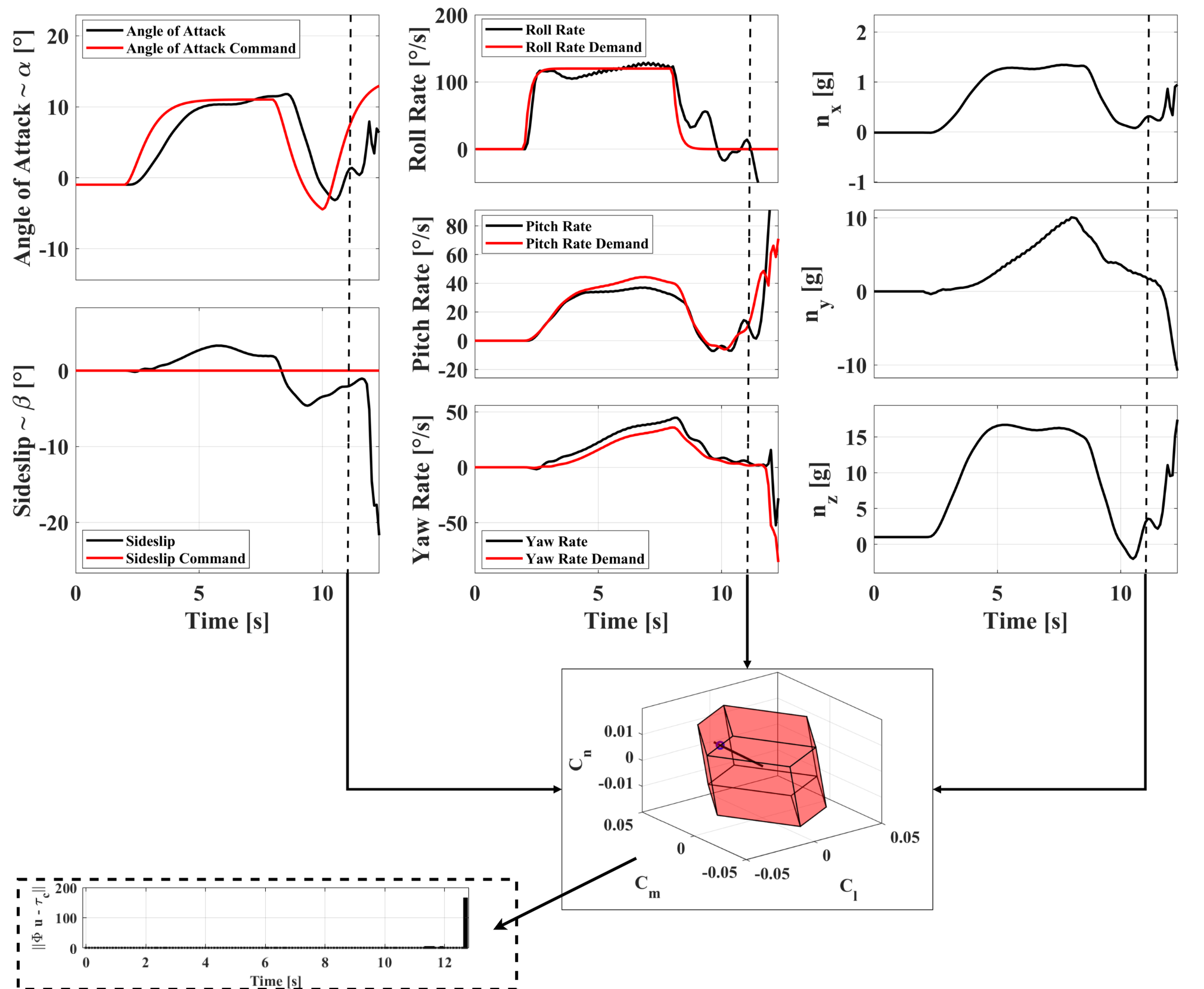}
\caption{State trajectory of the maneuver \#2 without loss of control prevention.}
\label{fig:statetrajectoryMan2woutLOCprevention}
\end{figure}

Initially, Fig.~\ref{fig:statetrajectoryMan2woutLOCprevention} depicts the maneuver performance of the aircraft without the loss of control prevention architecture.

%%%% without LOC prevention figures %%%%

\begin{figure}[hbt!]
\centering
\includegraphics[width=6.2in]{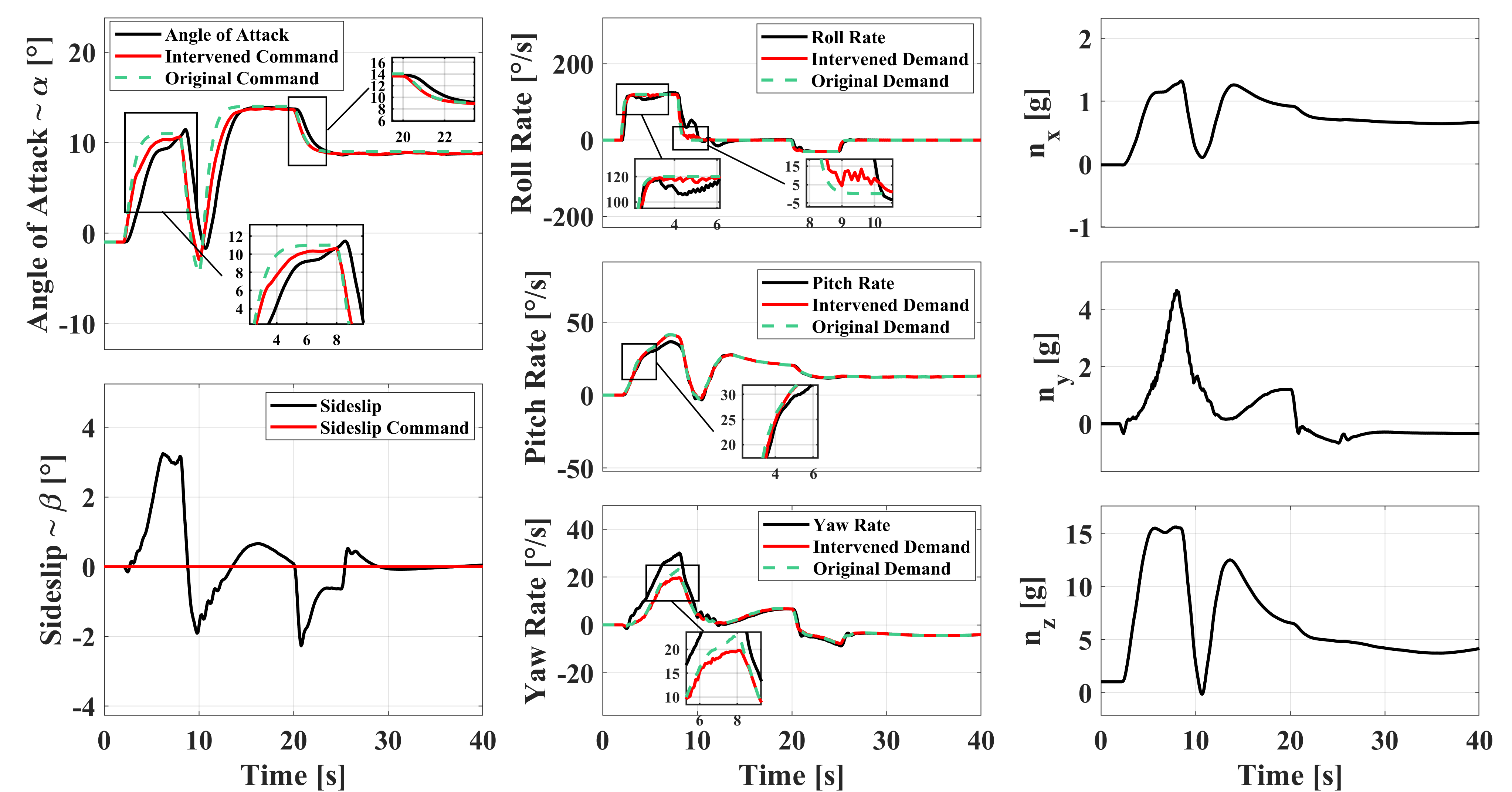}
\caption{State trajectory of the maneuver \#2 with loss of control prevention: indicates the intervention of the Lyapunov assessment.}
\label{fig:statetrajectoryMan2}
\end{figure}

Similar to the previous instance, the aircraft is incapable of executing the demanded maneuver due to the violation of the controllable space. Once again, violation of the controllability boundaries initiates a departure from controlled flight. The effects of the loss of control prevention architecture implementation are depicted in Fig.~\ref{fig:statetrajectoryMan2}.

%%%% with LOC prevention figures %%%%

\begin{figure}[hbt!]
\centering
\includegraphics[width=6in]{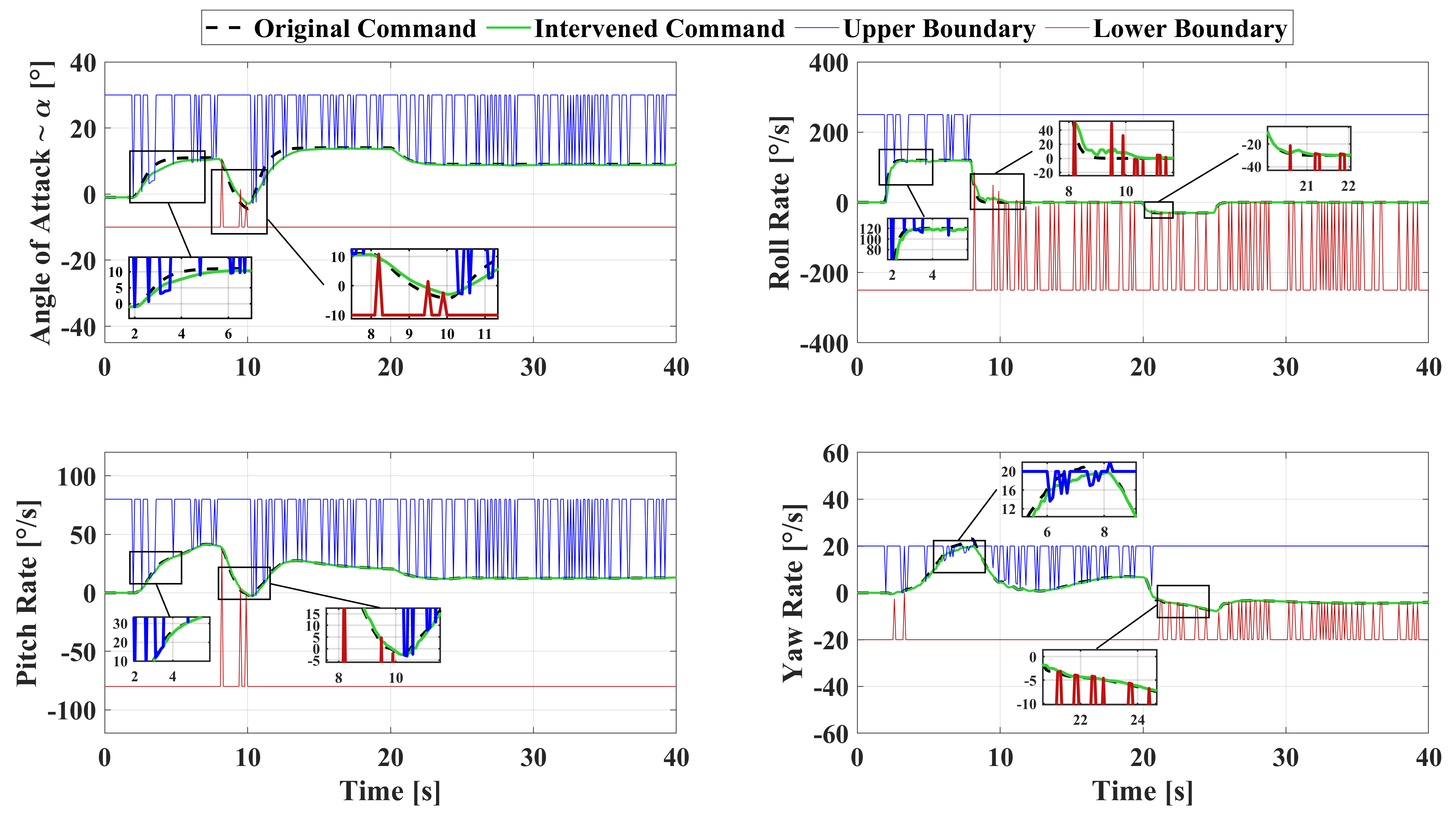}
\caption{Command saturation trajectory of the maneuver \#2 with loss of control prevention.}
\label{fig:boundarytrajectoryMan2}
\end{figure}

Furthermore, the variations in the saturation of the corresponding states of the aircraft are demonstrated in Fig.~\ref{fig:boundarytrajectoryMan2}.

\begin{figure}[hbt!]
  \centering
  \begin{tabular}[b]{c}
    \includegraphics[width=3.0in]{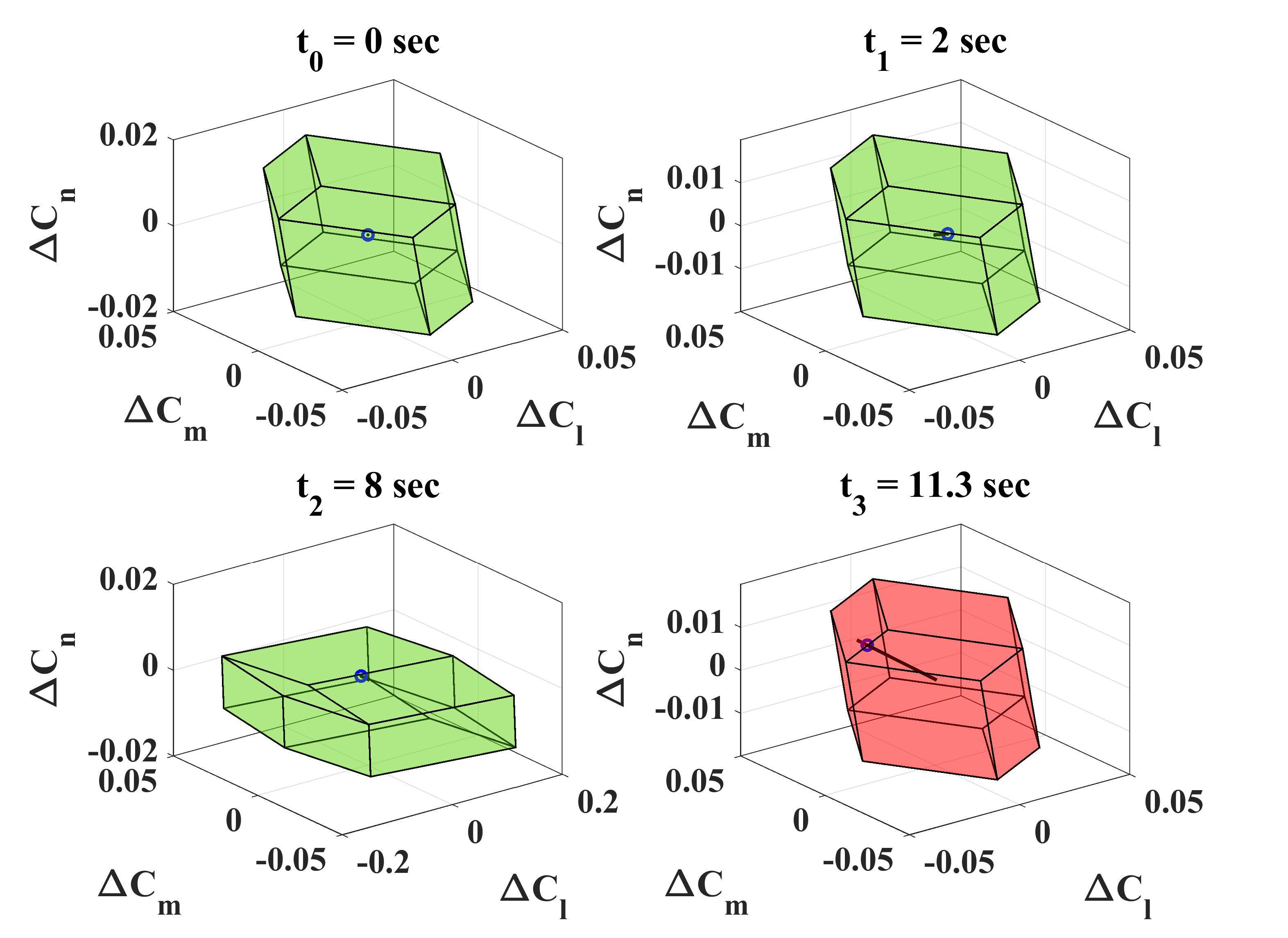} \\
    \small (a) Without loss of control prevention 
  \end{tabular} 
  \begin{tabular}[b]{c}
    \includegraphics[width=3.0in]{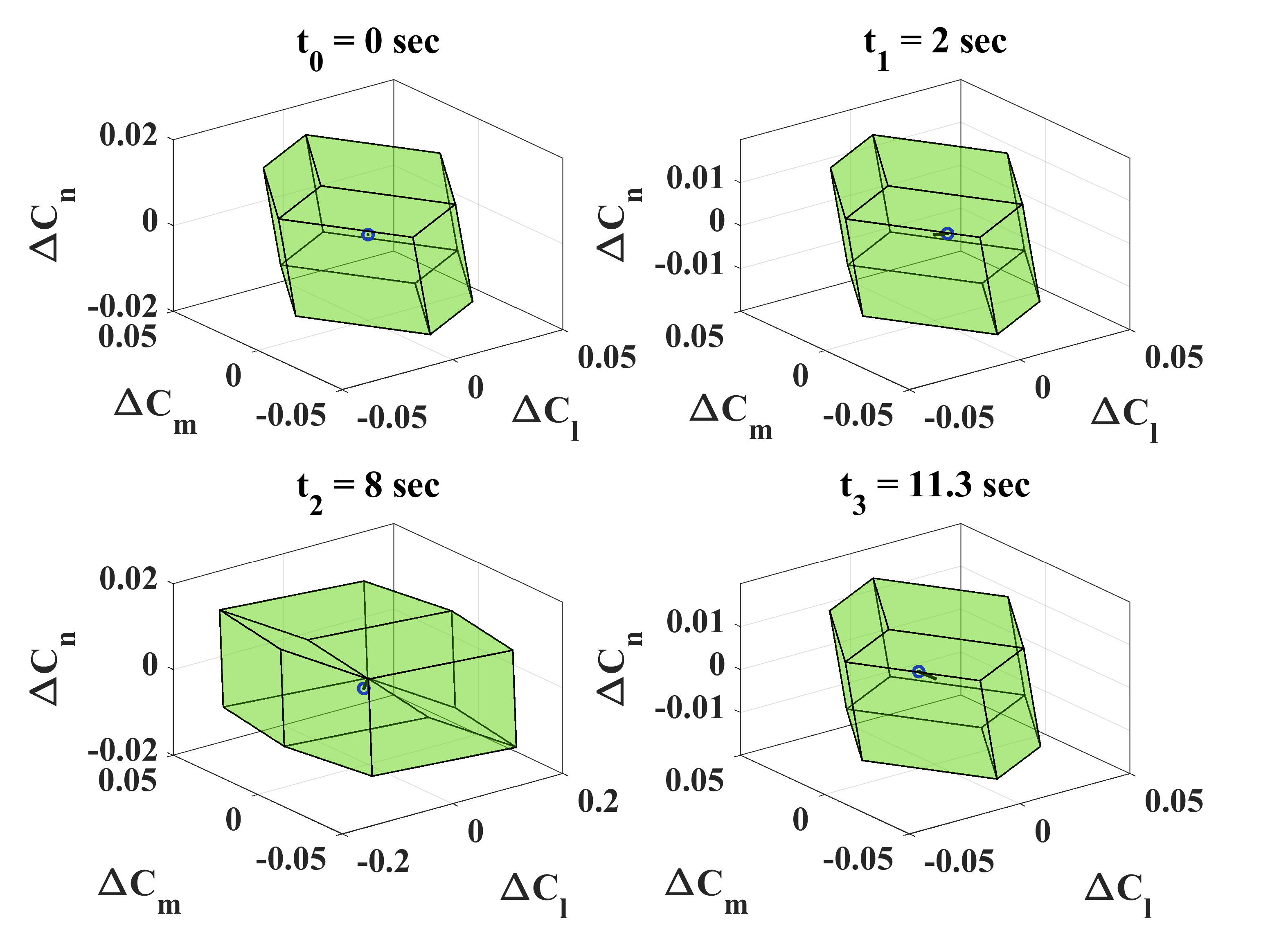} \\
    \small (b) With loss of control prevention
  \end{tabular} 
  \caption{Attainable moment set history: (a) without loss of control prevention, (b) with loss of control prevention during maneuver \#2.}
  \label{fig:man2_5}
\end{figure}

The attainable moment set histories are demonstrated in Fig.~\ref{fig:man2_5}.

\begin{figure}[hbt!]
 \centering
 \includegraphics[width=5.5in]{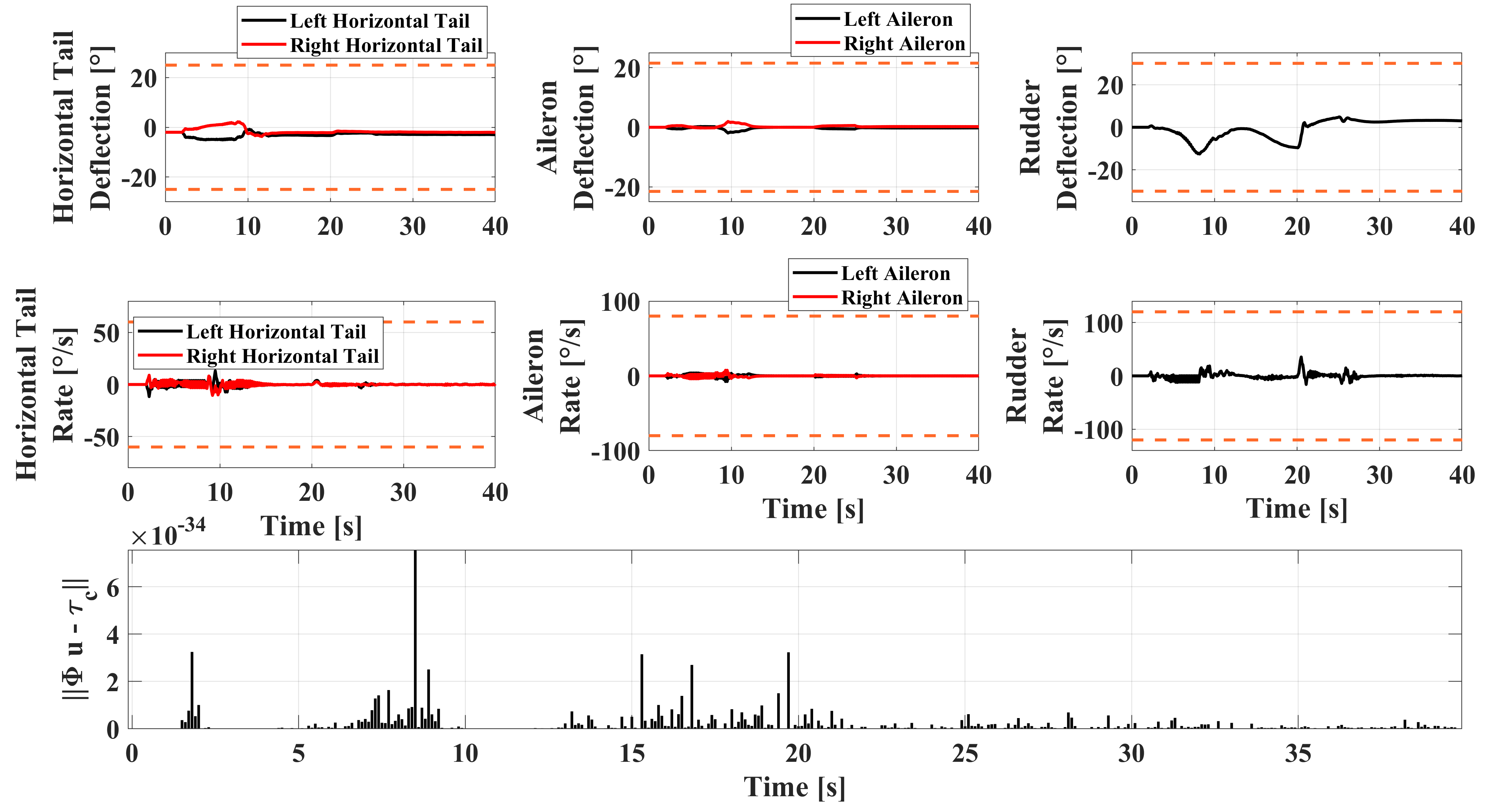}
 \caption{Control deflections with demanded moment satisfaction parameter of the maneuver \#2 with loss of control prevention.}
 \label{fig:controlMan2}
\end{figure}

\begin{figure}[hbt!]
 \centering
 \includegraphics[width=5.5in]{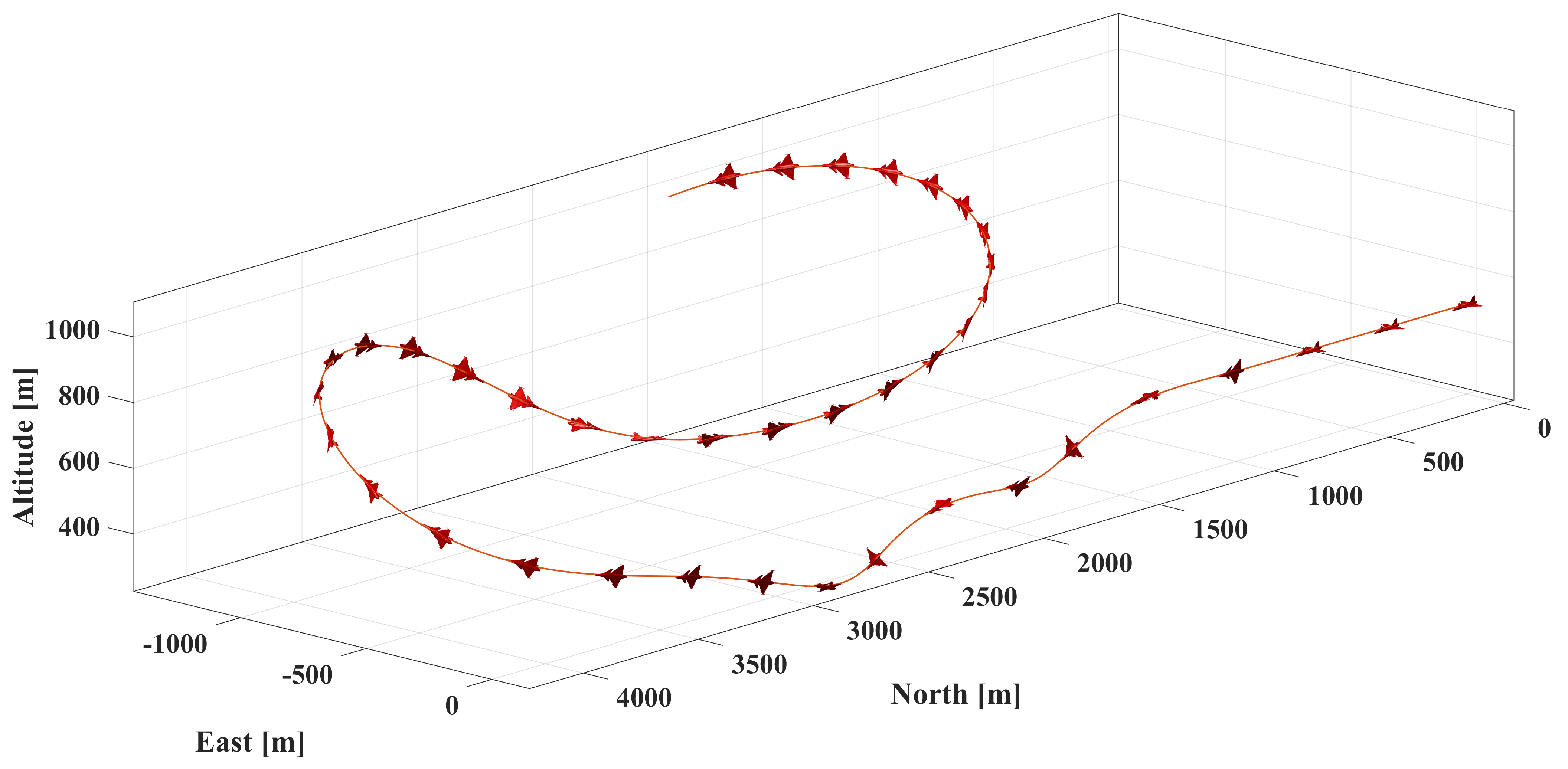}
 \caption{3D trajectory of the maneuver \#2 with loss of control prevention.}
 \label{fig:3DtrajMan2}
\end{figure}

Consequently, the history of the control deflections with demanded moment satisfaction parameter and the 3D trajectory of the maneuver are depicted in Fig.~\ref{fig:controlMan2} and Fig.~\ref{fig:3DtrajMan2} respectively.

Ultimately, without the loss of control prevention architecture, the aircraft is unable to execute the desired maneuver. It is proven that the aircraft outperforms this maneuver compared to the simulation without the loss of control prevention. It is also noteworthy that the maneuver is so extreme that the load factor approaches the limit of 15g.

\subsection{Comparison Study: Scheduled State Limiter versus Lyapunov-based Command Saturation}
\label{comparisonLyapunov}

\begin{figure}[hbt!]
 \centering
 \includegraphics[width=5in]{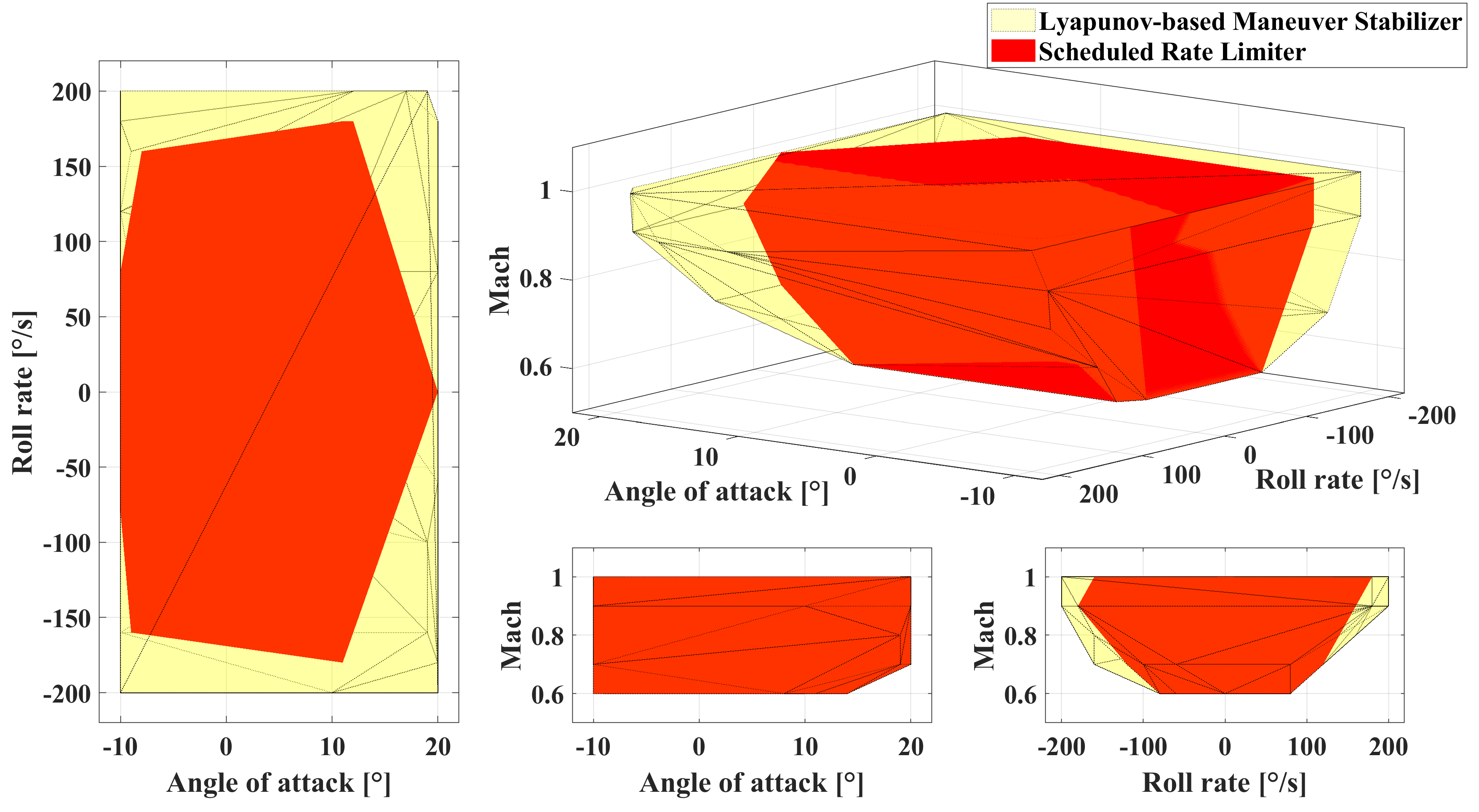}
 \caption{The comparison for sea level altitude: the yellow polygon's volume is 0.9504uc, while the red polygon's is 0.7111uc.}
 \label{fig:SLenvelope}
\end{figure}

\begin{figure}[hbt!]
 \centering
 \includegraphics[width=5in]{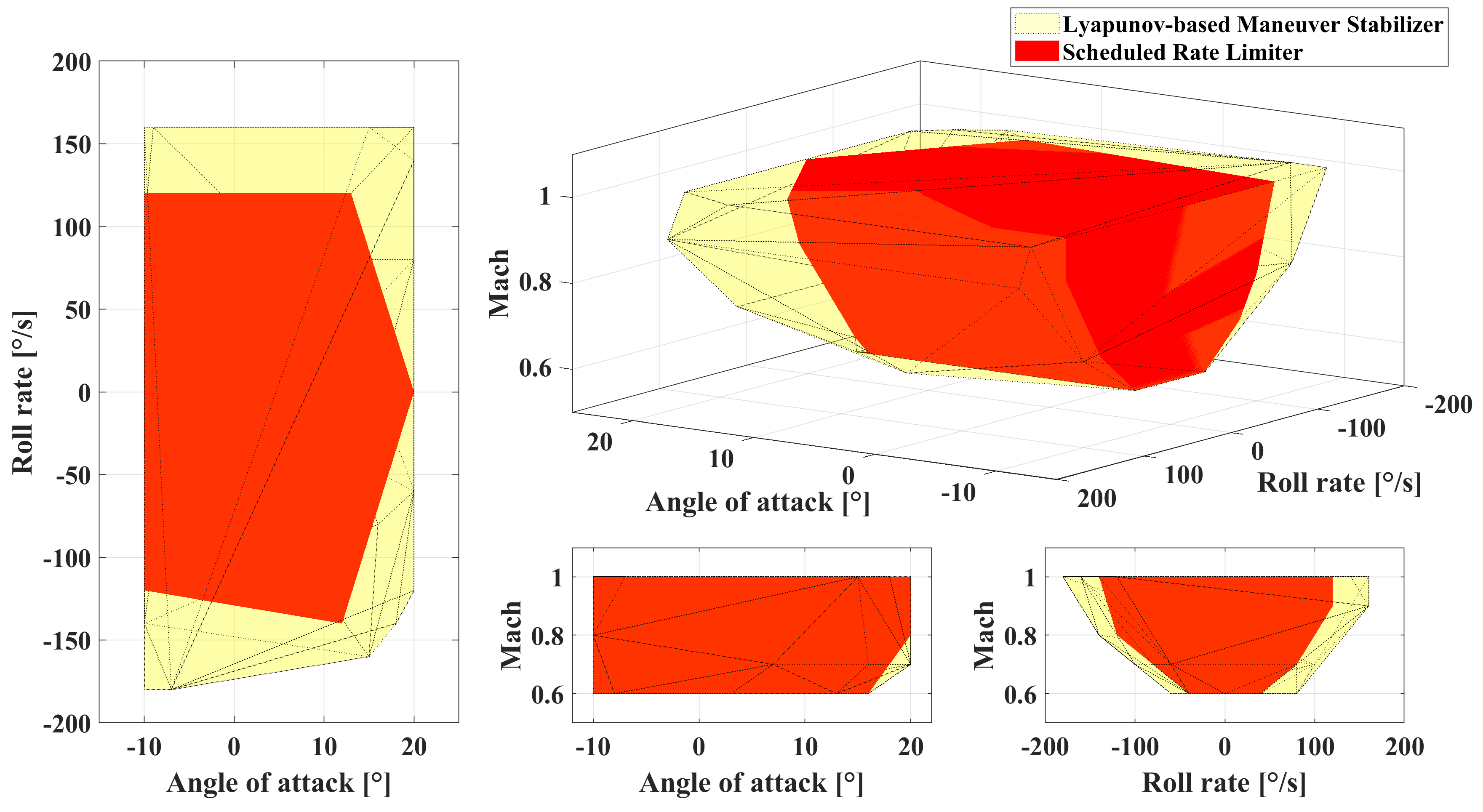}
 \caption{The comparison for 10000ft altitude: the yellow polygon's volume is 0.7907uc, while the red polygon's is 0.5094uc.}
 \label{fig:10000ftenvelope}
\end{figure}

The final assessment involves comparing the Lyapunov-based command saturation approach with the conventional scheduled state limiter. The limiters are designed for the angle of attack, roll rate, pitch rate, and yaw rate. To thoroughly investigate the proposed architecture, various initial conditions are evaluated. Firstly, the aircraft is trimmed for level flight at sea level and an altitude of 10000 feet by varying the Mach number within the interval of $[0.6 \hspace{0.15cm} 1]$ with an increment of 0.1. Then, at the \nth{1} second of the simulation, the coupling inputs of angle of attack and roll rate are applied to the system as step inputs while keeping the sideslip angle at $0^\circ$. Subsequently, the simulations are analyzed in terms of maintaining stable flight. For this assessment, Monte Carlo simulations are conducted for over 3000 cases for each altitude. The polygons shown include all stable points simulated, as given in Fig.~\ref{fig:SLenvelope} and Fig.~\ref{fig:10000ftenvelope}.

The most eye-catching distinction of these two separate assessment's properties is the reduction in the most agile and stable maneuver capability, which is highly related to the control authority alteration due to the available dynamic pressure. Furthermore, an observable distinction in achievable positive and negative roll rates exists because of the asymmetric properties of the aircraft, such as inertia coupling and the gyroscopic effect due to the engine rotating masses.

However, perhaps the most significant interpretation of these assessments is that the Lyapunov-based command saturation approach allows for more abrupt/higher amplitude maneuvers without compromising stable flight, which has been the main objective of the study. In terms of expansion, the Lyapunov-based command saturation approach increased the generated volume compared to the conventional approach by 33.66\% for the sea level altitude assessment and by 55.21\% for the 10000ft altitude assessment. The choice of using the expansion in the stable maneuverable volume as the assessment metric is fundamentally rooted in the objective of enabling higher amplitude maneuvers in a stable manner with the proposed approach, which aligns with the primary aim of this study. Therefore, this assessment serves as concrete evidence that more abrupt/higher amplitude and stable flight can be achieved using the proposed method. However, it is noteworthy that, even using the proposed method, not all combinations of angle of attack and roll rate maneuvers can be performed without losing control, probably reasoning from the lack of considering the dynamic pressure and inertial couplings in the loss of control detection through IAMS; additionally, the choice of Lyapunov gain $K$, which needs a further attention to be tuned.

\begin{comment}
    
\begin{table}[hbt!]
\begin{center}
\caption{Comparison}\label{tb:comparison}
\begin{tabular}{ccccccccccccccccccccc}
\hline
\hline
\multicolumn{20}{c}{Dynamic command saturation} \\
\multicolumn{10}{c}{Sea level} & & \multicolumn{10}{c}{10000ft} \\ 
\cline{1-10} & \cline{11-21}
\multicolumn{2}{c}{0.6M} & \multicolumn{2}{c}{0.7M} & \multicolumn{2}{c}{0.8M} & \multicolumn{2}{c}{0.9M} & \multicolumn{2}{c}{1M} & & \multicolumn{2}{c}{0.6M} & \multicolumn{2}{c}{0.7M} & \multicolumn{2}{c}{0.8M} & \multicolumn{2}{c}{0.9M} & \multicolumn{2}{c}{1M} \\
\hline
$\alpha$, $^\circ$ & $p$, $^\circ/s$ & $\alpha$, $^\circ$ & $p$, $^\circ/s$ & $\alpha$, $^\circ$ & $p$, $^\circ/s$ & $\alpha$, $^\circ$ & $p$, $^\circ/s$ & $\alpha$, $^\circ$ & $p$, $^\circ/s$ & $\alpha$, $^\circ$ & $p$, $^\circ/s$ & $\alpha$, $^\circ$ & $p$, $^\circ/s$ & $\alpha$, $^\circ$ & $p$, $^\circ/s$ & $\alpha$, $^\circ$ & $p$, $^\circ/s$ & $\alpha$, $^\circ$ & $p$, $^\circ/s$ \\
\hline

\hline
\hline
\end{tabular}
\end{center}
\end{table}

\end{comment}

%% 

\section{Conclusion \& Future Works}
\label{conc}

This study presents an online loss of control prevention method for an over-actuated F-16 model by combining Lyapunov's direct method with the aircraft's inverse dynamics. The flight control architecture integrates incremental nonlinear dynamic inversion and optimization-based control allocation, with pilot inputs being angle of attack ($\alpha$), sideslip angle ($\beta$), and roll rate ($p$). The challenge lies not only in controlling the aircraft but also in ensuring flight safety by eliminating destabilizing effects. To address this critical issue, a structure with two components is developed: a controllability investigation module (using incremental attainable moment set) and a online command saturation module (merging Lyapunov's direct method and inverse dynamics of the aircraft). Through the proposed structure, the control commands are saturated during flight to prevent loss of control in the event of a detection of a loss of control possibility. Prior to applying online saturation, a controllability assessment is conducted using an incremental attainable moment set, which considers actuator limitations such as position and rate saturation. This assessment determines whether the demanded control moment vector falls within the attainable set or not, whereas the allowable control authority is set at 70\%, with a 30\% safety margin. In accordance with the assumptions and the scope of the study, violating this set indicates an excess of control authority, and in such cases, the Lyapunov-based dynamic command saturation is activated. The proposed architecture's effectiveness is demonstrated through two challenging maneuvers, showing that stable maneuvers can be maintained only with the loss of control prevention structure. A comparison with conventional state limiters reveals a significant expansion in the stable maneuverable volume (33.66\% at sea level and 55.21\% at 10000 ft). The architecture extends the aircraft's safe maneuvering envelope, enabling abrupt and agile maneuvers without compromising stability. Further studies are planned to examine the effects of the scheduled gain on the expansion of the stable maneuverable volume using an adaptive Lyapunov gain. Additionally, another aim is to explore the compatibility of the presented method with different aircraft types, such as commercial aviation airplanes or quadcopters. Hence, the plan includes working on the real-time implementation of various baselines and investigating scenarios involving a fault occurrence, utilizing either the presented methodology or an improved version, as a part of future work.

\bibliography{mybibfile}

\end{document}